\RequirePackage{snapshot}
\documentclass[aps,prb,twocolumn,amsfonts,showpacs,longbibliography,superscriptaddress,floatfix]{revtex4-2}
\usepackage{verbatim,epsfig,amsmath,amssymb,bm,epsf,graphicx,psfrag,bbold,amsthm,amsfonts}
\usepackage[bottom]{footmisc}
\usepackage{hyperref,array}
\usepackage[capitalize]{cleveref} 
\usepackage{enumitem}
\usepackage{grffile}
\usepackage{framed}
\usepackage{mathrsfs}
\usepackage{empheq}
\usepackage{esint}
\setlist{nosep}
\usepackage{placeins}
\usepackage[all]{xy}
\usepackage{color}
\usepackage[utf8]{inputenc}
\usepackage{float}
\usepackage{natbib}
\usepackage{tikz-cd}
\usepackage{verbatim}
\usepackage{leftidx}
\usepackage[normalem]{ulem}

\usepackage{notes2bib}
\newcommand{\moy}[1]{\langle #1 \rangle}

\newcommand{\longsquiggly}{\xymatrix{{}\ar@{~>}[r]&{}}}

\newcommand{\cM}{\mathcal{M}}

\newcommand\beq {\begin{equation}}
	\newcommand\eeq {\end{equation}}
\newcommand\beqa {\begin{equatiobn}\begin{array}}
		\newcommand\eeqa {\end{array}\end{equation}}
\newcommand\bal {\begin{align}}
	\newcommand\eal {\end{align}}
\newcommand{\bea}{\begin{eqnarray}}
	\newcommand{\eea}{\end{eqnarray}}

\theoremstyle{plain}

\theoremstyle{definition}

\theoremstyle{remark}

\begin{document}
	
	\title{Classical Non-Relativistic Fractons}
	\author{Abhishodh Prakash}
	\email{abhishodh.prakash@physics.ox.ac.uk (he/him/his)}
	\affiliation{Rudolf Peierls center for Theoretical Physics, University of Oxford, Oxford OX1 3PU, United Kingdom}
	
	\author{Alain Goriely}
	\email{alain.goriely@maths.ox.ac.uk}
	\affiliation{Mathematical Institute, University of Oxford, Oxford OX2 6GG, United Kingdom}
	
	\author{S.L. Sondhi}
	\email{shivaji.sondhi@physics.ox.ac.uk}
	\affiliation{Rudolf Peierls center for Theoretical Physics, University of Oxford, Oxford OX1 3PU, United Kingdom}


	\begin{abstract}
		We initiate the study of the classical mechanics of non-relativistic
fractons in its simplest setting - that of identical one-dimensional
particles
with local Hamiltonians characterized by a conserved dipole moment in
addition to the usual symmetries of space and time translation invariance.
We introduce a family of models and study the $N$-body problem for them.
We find that locality leads to a ``Machian" dynamics in which a given
particle exhibits finite inertia only if within a specified distance of another particle. For well-separated particles, this dynamics leads to
immobility,
much as for quantum models of fractons discussed before. For two or more
particles within inertial reach of each other at the start
of motion we obtain an interesting interplay of inertia and interactions.
Specifically, for a solvable ``inertia only" model of fractons, we find that two particles always become immobile at long times. Remarkably, three
particles generically evolve to a late time state with one immobile particle and two oscillating about a common center of mass with generalizations of
such ``Machian clusters" for  $N > 3$ . Interestingly, these Machian clusters exhibit
physical limit cycles in a Hamiltonian system even though mathematical limit cycles are forbidden by Liouville's theorem.
	\end{abstract}
	
	\maketitle
	
	\tableofcontents
	\newpage 
	\section{Introduction and summary of main results}
 
Much recent work in quantum many-body theory has focused on the properties of so-called ``fracton" or ``fractonic" phases of matter. These phases~\cite{NandkishoreHermeleFractonsannurev-conmatphys-031218-013604,PretkoChenYou_2020fracton,GromovRadzihovsky2022fractonReview} host excitations with restricted mobility of which the ones that are immobile in isolation are termed fractons. This term in the current context is a legacy of the paper that kicked off the boom in which Haah discovered an exceptionally complicated example \cite{Haah_FractonPhysRevA.83.042330} wherein the operators that create widely separated fractons are fractal in character\footnote{Once the term fractons was reserved for excitations on a fractal background \href{https://doi.org/10.1016/0167-2789(89)90204-2}{sciencedirect.com/science/article/pii/0167278989902042} but that copyright expired a while back.}. Subsequently, it was realized that the basic phenomenon of immobility did not require fractality but could be obtained with simpler membrane operators as shown by Chamon  before~\cite{Chamon_Fracton_PhysRevLett.94.040402}. A second important simplification was the realization that full-blown fractons could be obtained by gauging models with unusual symmetries, which already exhibited restricted particle mobility. The simplest such ungauged models exhibit the conservation of higher moments of the charge distribution i.e. multipoles and recent work has explored the effects of multipole conservation in continuum and many-body quantum mechanical settings\cite{NandkishoreHermeleFractonsannurev-conmatphys-031218-013604,PretkoChenYou_2020fracton,GromovRadzihovsky2022fractonReview}.

Here, we initiate the study of {\it classical}, non-relativistic, systems of point particles with multipole symmetries in the continuum. We do this by studying the simplest possible system with a conserved dipole moment in $d=1$ spatial dimension. For such systems we study its Hamiltonian dynamics for a fixed number $N$ of point particles. 

Despite its simplicity this system exhibits many surprises that defy conventional intuition about classical particle systems. Overall we find that symmetry and locality conspire to produce a variant of Machian dynamics~\cite{Mach1907science,Mach_bookbarbour1995mach,BondiSamuel_Machj1997121,Pretko_MachPhysRevD.96.024051} wherein particles exhibit inertia only in the presence of others and are motionless in isolation. The nature of dynamics exhibited by Machian clusters depends on the number of particles in close proximity, in addition to the nature of the Hamiltonian and initial conditions. In the absence of additional particle interactions, $N=2$ particles in close initial separation generically separate and become immobile with the passage of time. In the case $N=3$, particles generically settle down to a steady-state dynamics in position space where one particle becomes immobile and the other two oscillate about a center of mass in the manner of a limit cycle, seemingly violating (but ultimately consistent with) Liouville's theorem. An exact solution for $N=3$ particles is obtained in an idealized limit. For $N>3$, numerical integration of the equations of motion results in the particles separating into at least two clusters  exhibiting motionless, oscillatory, or chaotic dynamics at late times. Other unusual features that characterize the trajectories are (i) the breakdown of the customary relationships between momenta, velocity, and energy, (ii) constant energy phase-space hypersurfaces that are unbounded in the momenta, and (iii) the appearance of asymptotic, emergent, conserved quantities. Throughout the paper, we will refer to dipole-conserving particles as `fractons' which is now standard practice~\cite{SkinnerPozderac_Thermalization_2023,KhemaniHermeleNandkishore_Shattering_PhysRevB.101.174204,SalaRakovskyVerresenKnapPollmann_FragmentationPhysRevX.10.011047}.
		
    The paper is organized as follows. In \cref{sec:Symmetries_and_compatible_hamiltonians}, we review how symmetries are implemented in classical Hamiltonian systems and write down the form of Hamiltonians compatible with dipole symmetry. In \cref{sec:Classical trajectories}, we numerically study the classical trajectories of $N=1,2,3,4$ fractons and in \cref{sec:exact_trajectories}, we obtain the exact solution for the trajectories for $N=2,3$ for a specific form of the Hamiltonian. In \cref{sec:Discussions}, we discuss specific aspects of fracton trajectories and parallels with known quantum fracton phenomenology in \cref{sec:Quantum} before concluding. 
	
	\section{Symmetries and compatible Hamiltonians}
	\label{sec:Symmetries_and_compatible_hamiltonians}
	\subsection{Symmetries in classical non-relativistic systems of point-particles}
	
	Consider a system of $N$ point particles living in $d$ spatial dimensions. We will work within a classical Hamiltonian framework to describe its dynamics. Here, the state of the system is specified by $2dN$  phase space coordinates consisting of $dN$ positions $\{x^\mu_{1},\ldots, x^\mu_{N}\}$ and $dN$ momenta $\{p^\mu_{1} \ldots p^\mu_{N}\}$. The dynamics is generated by  the Hamiltonian $H(\{x^\mu_a,p^\mu_a\})$ through the usual Hamilton's equations of motion:
	\begin{equation}
		\dot{x}^\mu_a = \frac{\partial H}{\partial p^\mu_a},~\dot{p}^\mu_a = -\frac{\partial H}{\partial x^\mu_a}. \label{eq:Hamilton_eom}
	\end{equation}
	Above and henceforth, we will use the short hand notation $\dot{A} \equiv \frac{\partial A}{\partial t}$, the superscripts $\mu=1,\ldots,d$ to denote the spatial components of the phase space coordinates and subscripts $a=1,\ldots,N$ to specify each particle. We will also sometimes vectorize the spatial components of phase-space coordinates as $\vec{x}_a,~\vec{p}_a$ etc.  
	
	Our main interest is to understand how Hamiltonian dynamics is constrained by dipole symmetries. As a warm-up, we begin by considering the conservation of total charge.   Assuming the $N$ particles carry electric charges $\{q_1,\ldots,q_N\}$,  the charge density $\rho$ and current $\vec{j}$, given by,
	\begin{align}
		\rho(x) &\equiv \sum_{a=1}^N q_a \delta^d(\vec{x}_a - \vec{x}), \\
		\vec{j}(x) &\equiv \sum_{a=1}^N q_a \dot{\vec{x}}_a \delta^d(\vec{x}_a - \vec{x}),
	\end{align}
	satisfy the continuity equation 
	\begin{equation}
		\dot{\rho} + \vec{\nabla}.\vec{j} =0, \label{eq:continuity}
	\end{equation}
	leading to the conservation of total charge $Q$
	\begin{equation}
		\dot{Q} =0~\text{ where, }Q \equiv \int d^dx ~\rho(x) = \sum_a q_a. \label{eq:Q_definition}
	\end{equation}
	Note that this conservation is a kinematic constraint and is built into the kinematic framework of classical non-relativistic Hamiltonian dynamics since the number of particles $N$ is fixed and cannot change. Consequently, ~\cref{eq:continuity,eq:Q_definition} hold regardless of Hamiltonian form. This is not true of other symmetries which constrain the form of the Hamiltonian. For example, if we want to impose translation symmetry generated by
	\begin{equation}
		x^\mu_a \mapsto x^\mu_a + \xi^\mu,~p^\mu_a \mapsto p^\mu_a, \label{eq:translation_CM}
	\end{equation}
	the form of the Hamiltonian is constrained to be
	\begin{equation}
		H(x^\mu_a,p^\mu_a)  \equiv H \left(  x^\mu_a - x^\mu_b , p^\mu_a \right) \label{eq:translation_Hamiltonian}
	\end{equation}
	and leads to the conservation of the total momentum
	\begin{equation}
		\dot{P}^\mu = 0, \text{where } ~P^\mu \equiv \sum_{a=1}^N p^\mu_a.
	\end{equation}
	We now turn to the main interest of this work and look to impose the conservation of the total dipole moment,
	\begin{align}
		\dot{D}^\mu=0,~D^\mu&\equiv \int d^dx~x^\mu\rho(x) = \sum_{a=1}^N q_a x^\mu_a  \label{eq:dipole_CM}
	\end{align}
	which induces the following symmetry transformation on the phase space coordinates
	\begin{equation}
		x^\mu_a \mapsto x^\mu_a ,~p^\mu_a \mapsto p^\mu_a + q_a \phi^\mu. \label{eq:dipolesymmetry_CM}
	\end{equation}
	Comparing \cref{eq:translation_CM,eq:dipolesymmetry_CM}, we see that dipole conservation implies translation in momentum space. This property constrains the Hamiltonian to be of the form
	\begin{equation}
		H(x^\mu_a,p^\mu_a) \equiv H\left(x^\mu_a,\frac{p^\mu_a}{q_a}-\frac{p^\mu_b}{q_b}\right).  \label{eq:dipole_Hamiltonian}
	\end{equation}		
	This type of construction can be easily  generalized for other constraints. Indeed,  $Q$ in \cref{eq:Q_definition} is the zeroth moment of charge distribution, whereas $\vec{D}$  in \cref{eq:dipole_CM}  is the first moment. Therefore, we can further consider the  conservation of arbitrary combinations of charge moments such as
	\begin{equation}
		\dot{M}_P =0,	M_P = \int d^dx ~P(x)  \rho(x) = \sum_a q_a P(x_a)
	\end{equation} 
	where $P(x)$ is a polynomial in the $d$ components $x^\mu$. This constraint induces a so-called `polynomial shift' symmetry
	\begin{equation}
		x^\mu_a \mapsto x^\mu_a ,~p^\mu_a \mapsto p^\mu_a + q_a \lambda \frac{\partial P(x_a)}{\partial x^\mu_a}. \label{eq:polynomial_CM}
	\end{equation}	
	and generates the so-called multipole algebra discussed in Ref.~\cite{Gromov_Multipole_PhysRevX.9.031035} of which \cref{eq:dipolesymmetry_CM} is a special case.
	Then, symmetry compatible	Hamiltonians which obey  \cref{eq:polynomial_CM} are constrained to take the form 	
	\begin{equation}
		H \rightarrow H\left(\frac{p^\mu_a}{q_a} \left(\frac{\partial P(x_a)}{\partial x^\mu_a}\right)^{-1}-\frac{p^\mu_b}{q_b} \left(\frac{\partial P(x_b)}{\partial x^\mu_b}\right)^{-1}\right),\label{eq:multipole_Hamiltonian}
	\end{equation}
	where repeated indices are not summed over.

 For the rest of this paper, we will restrict ourselves to dipole conservation symmetry shown in \cref{eq:dipole_CM} and study the dynamics generated by Hamiltonians of the form \cref{eq:dipole_CM} and leave the more general case of \cref{eq:polynomial_CM} for future work.
	
	\subsection{Symmetry compatible Hamiltonians}
	\label{sec:Symmetry_Hamiltonian}
	We now focus on the dipole conservation symmetry of \cref{eq:dipole_CM,eq:dipolesymmetry_CM} and look at the possible forms the Hamiltonian of \cref{eq:dipole_Hamiltonian} can take. Recall that translation invariance of coordinates shown in \cref{eq:translation_CM} results in the Hamiltonian depending only on the difference of the coordinates as shown in \cref{eq:translation_Hamiltonian}.  A common form of such a Hamiltonian is
	\begin{equation}
		H = \sum_{a} \frac{|\vec{p}_a|^2}{2 m_a} + \sum_{a<b} U(x_a - x_b) + \ldots \label{eq:Hamiltonian_translation_explicit}
	\end{equation} 
	In order to be physically reasonable, we require the Hamiltonian to be  \emph{local}, i.e., particles with large distances $|\vec{x}_a - \vec{x}_b|$ between them should not affect each others dynamics. This requires the  two-particle interaction in \cref{eq:Hamiltonian_translation_explicit}, represented by $U(x)$ to vanish as $|x| \rightarrow \infty$. Also, the quadratic form of the kinetic term, $|\vec{p}_a|^2$ ensures that energies are bounded from below. Let us now move on to the case of a Hamiltonian with dipole symmetry which has a functional form shown in \cref{eq:dipole_Hamiltonian}. A natural modification of \cref{eq:Hamiltonian_translation_explicit} to accommodate this is
	\begin{equation}
		H =\sum_{a<b} \left(\frac{1}{2 m_{ab}}  \left|\frac{\vec{p}_a}{q_a}-\frac{\vec{p}_b}{q_b}\right|^2 +  U(x_a - x_b) + \ldots\right) \label{eq:Hamiltonian_dipole_nonlocal}
	\end{equation} 
	However, note that the leading momentum-dependent kinetic term is no longer local. To restore locality, we  modify it as follows
	\begin{equation}
		H = \sum_{a<b}\left(\frac{K(|\vec{x}_a - \vec{x}_b|)}{2 m_{ab}}  \left|\frac{\vec{p}_a}{q_a}-\frac{\vec{p}_b}{q_b}\right|^2  +  U(x_a - x_b) + \ldots \right),\label{eq:Hamiltonian_dipole_explicit}
	\end{equation} 
	where $K(x)$ is some function with local support that vanishes as $|x| \rightarrow \infty$ to impose locality. For the energies to be bounded from below,  $K(x)$ should also be positive definite. 
	
	\subsection{Mach's principle}
	\label{sec:Mach}
	
	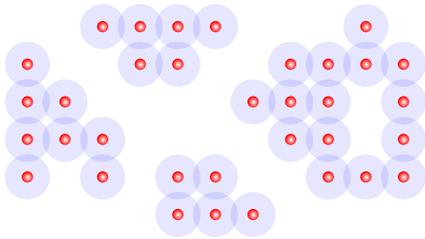
\begin{figure}[!ht]
		\begin{tikzpicture}[scale=.5]
			
			\foreach \x in {5,6,7} 
			{\fill[blue,opacity=0.1] (\x ,-1) circle (0.6);
				\shade[inner color=white, outer color = red] (\x ,-1) circle (0.15);}
			
			\foreach \x in {1,3,5,6,9,10,11} 
			{\fill[blue,opacity=0.1] (\x ,0) circle (0.6);
				\shade[inner color=white, outer color = red] (\x ,0) circle (0.15);}
			
			\foreach \x in {1,2,3,8,9,11} 
			{\fill[blue,opacity=0.1] (\x ,1) circle (0.6);
				\shade[inner color=white, outer color = red] (\x ,1) circle (0.15);}
			
			\foreach \x in {1,2,7,8,9,11} 
			{\fill[blue,opacity=0.1] (\x ,2) circle (0.6);
				\shade[inner color=white, outer color = red] (\x ,2) circle (0.15);}
			
			\foreach \x in {1,4,5,8,9,10,11} 
			{\fill[blue,opacity=0.1] (\x ,3) circle (0.6);
				\shade[inner color=white, outer color = red] (\x ,3) circle (0.15);}
			
			\foreach \x in {3,4,5,6,10} 
			{\fill[blue,opacity=0.1] (\x ,4) circle (0.6);
				\shade[inner color=white, outer color = red] (\x ,4) circle (0.15);}
		\end{tikzpicture}
		\caption{Sample particle configuration in d=2 corresponding to four visible Machian clusters. The dark red circles represent particles and the shaded blue region surrounding them represent the range where $K \neq 0$.   \label{fig:Mach}}
	\end{figure}
	
	The unusual form of the kinetic term in \cref{eq:Hamiltonian_dipole_explicit} has several consequences which we will explore in the upcoming sections. We point out one of them here -- the familiar relation between momentum and velocity -- $\dot{x}_a \propto p_a$ no longer holds for \cref{eq:Hamiltonian_dipole_explicit}. Instead,  we have
	\begin{equation}
		\dot{x}^\mu_a = \frac{\partial H}{\partial p^\mu_a} = \sum_b \left(\frac{p^\mu_a}{q_a} - \frac{p^\mu_b}{q_b}\right)  \frac{ K(|x^\mu_a - x^\mu_b|)}{q_a m_{ab}}. \label{eq:velocity_dipole}
	\end{equation}
	From the form of the velocities in \cref{eq:velocity_dipole}, we can directly obtained dipole conservation:
	\begin{equation}
		\dot{D}^\mu = \sum_a q_a \dot{x}^\mu_a = \sum_{a,b} \left(\frac{p^\mu_a}{q_a} - \frac{p^\mu_b}{q_b}\right)  \frac{ K(|x^\mu_a - x^\mu_b|)}{ m_{ab}} =0. \nonumber
	\end{equation}
	
	\Cref{eq:velocity_dipole} also  tells us that the velocity of a given particle is affected by all others. This is reminiscent of Mach's principle~\cite{Mach1907science} which states that motion of a body is not absolute but is influenced by all others in the cluster. In the context of gravity where this principle has been most discussed~\cite{Mach_bookbarbour1995mach,BondiSamuel_Machj1997121}, the formulation aims to describe the motion of particles within a single cluster and is inherently non-local. For instance, the kinetic term in Ref.~\cite{Lynden-Bell_Katz_MachianCM_PhysRevD.52.7322} is similar to the one shown in \cref{eq:Hamiltonian_dipole_nonlocal}. 
 
 The inclusion of the locality term $K(x)$ which is natural in the context of condensed matter physics changes the picture qualitatively. Now, the motion of a particle, located at position $\vec{x}_a$ is influenced, not by all  particles, but only those  particles located at $\vec{x}_b$ such that $K(\vec{x}_a - \vec{x}_b) \neq 0$. The dynamics of \cref{eq:Hamiltonian_dipole_explicit} is therefore that of particles living in and moving between multiple dynamical clusters as shown in \cref{fig:Mach}. We use the term `Machian cluster' to refer to a collection of particles which lend inertia and influence each others motion. Notice that the number of Machian clusters can change under dynamics as particles move apart or come together.  The connection between dipole conserving systems and Mach's principle was originally pointed out in Ref.~\cite{Pretko_FractonGauge_PhysRevB.98.115134} using a continuum field theoretic formulation, for quantum systems. Our non-relativistic classical formulation makes this connection much more transparent.

	Note that \cref{eq:velocity_dipole} cannot be inverted. In other words, given a collection of velocities $\{\dot{x}^\mu_a\}$, we do not obtain a unique set of momenta $\{p^\mu_a\}$. This is a consequence of dipole conservation-- observe that the expression for velocities $\{\dot{x}^\mu_a\}$ in \cref{eq:velocity_dipole} is invariant under the momentum shift symmetry generated by dipole moment shown in \cref{eq:dipolesymmetry_CM}. An important consequence of this property is that momenta can reach  large values even when the velocities and energies are small. As we will see, this fact is at the heart of the novel physics present in these dipole-conserving systems.

	\section{Identical fractons on a line}
	\label{sec:Classical trajectories}	
	\subsection{Preliminaries and general features}

	\subsubsection{Preliminaries}

\begin{figure}[!htbp]
		\centering
		\begin{tabular}{c}
			\includegraphics[width=0.4\textwidth]{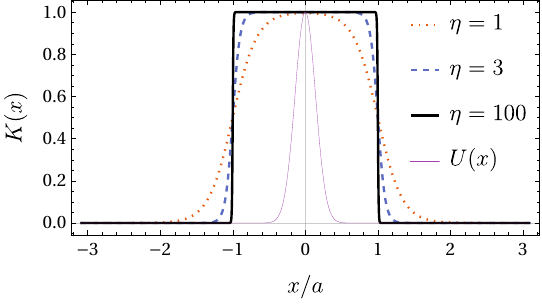}  
		\end{tabular}
		\caption{$K(x)$ defined in \cref{eq:K(x) regulated} for various values of $\eta$ compared with $U(x)$ defined in \cref{eq:U(x)} with $\Gamma = 1$ and $b = 0.2$ . As $\eta \rightarrow \infty$, $K(x)$ takes the form shown in \cref{eq:K(x) box}.}
		\label{fig:K}
	\end{figure}

	Let us consider now the classical trajectories of  dipole-conserving systems in one dimension with identical charge and mass (both of which we set to unity without  loss of generality).  The dipole moment $D = \sum_a x_a$ is then proportional to the center-of-mass (COM). This is perfectly dual to the conservation of the center of momentum imposed by translation invariance: 
	\begin{align}
		\text{Translation: }& 		x_a \mapsto x_a + \xi ,~ P = \sum_a p_a \text{ conserved.} \nonumber\\
		\text{Dipole: }& 		p_a \mapsto p_a + \phi ,~ D = \sum_a x_a \text{ conserved.} \nonumber
	\end{align}
	The Hamiltonian  (\ref{eq:Hamiltonian_dipole_explicit}) now reads:
	\begin{equation}
		H = \sum_{a<b=1}^N \left[\frac{1}{2} \left(p_{a} - p_b \right)^2 K(x_{a} - x_b)    + U(x_{a} - x_b)\right]. \label{eq:Hamiltonian_classical_1d}
	\end{equation} 	
	For concreteness, we will use the following forms of $K(x)$ and $U(x)$ 
	\begin{align}
		K(x) &= \frac{1}{2} \left[\tanh\left(\eta \left(x+ a\right)\right)-\tanh\left(\eta \left(x- a\right)\right)\right] \label{eq:K(x) regulated},  \\
		U(x) &= \Gamma \exp \left(- \frac{x^2}{b^2}\right). \label{eq:U(x)}
	\end{align}
	although we expect all our results to hold for any other physically sensible choices of these functions. Here, $\Gamma$ denotes the interaction strength which can be attractive or repulsive, and  $\Gamma = 0$ represents the closest equivalent to free particles. The function $K(x)$ is  positive definite function and vanishes at $|x| \rightarrow \infty$. Its limiting form is a box function with compact support:
	\begin{equation}
		\lim\limits_{\eta \rightarrow \infty}	K(x) \equiv B(x) = \Theta(x+a) - \Theta(x-a), \label{eq:K(x) box}
	\end{equation}
	where $\Theta(x)$ is the Heaviside Theta function. Examples of these function is given in \cref{fig:K}. 
 
 We now have two length-scales in the Hamiltonian (\ref{eq:Hamiltonian_classical_1d}):  $a$ represents the separation scale below which the kinetic term is operative and two fractons can `lend inertia' to each other and $b$ represents the interaction range between particles. We will only consider the case where $a>b$ in what follows although we find the results qualitatively unchanged for $a<b$. 
	
	\subsubsection{Far-separated particles and general features}
	\begin{figure}[!ht]
		\begin{tikzpicture}[scale=.6]
			
			\foreach \x in {0,2,5,9,11} 
			{\fill[rounded corners,blue,opacity=0.1] (\x-0.8,-0.3) rectangle (\x+0.8,0.3);
				\shade[inner color=white, outer color = red] (\x ,0) circle (0.15);}
			
		\end{tikzpicture}
		\caption{Sample particle configuration in d=1 where initially all particle are sufficiently separated from each other and remain immobile. \label{fig:1d_farseparated}}
	\end{figure}
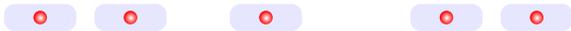
	\label{sec:far_sparated}
	We start with a configuration that immediately exposes general features of the dynamics generated by    \cref{eq:Hamiltonian_classical_1d} by switching off the interactions ($U(x)=0$) and set $\eta \rightarrow \infty$ when $K(x)$ takes the form  (\ref{eq:K(x) box}). For initial conditions when particles are well-separated $(x_a(0) - x_b(0)) \gg a$ as shown in \cref{fig:1d_farseparated},  the Hamiltonian (\ref{eq:Hamiltonian_classical_1d}) vanishes and all particles are  frozen in place. Thus, we have a large manifold of zero-energy configurations where each particle is isolated within an independent Machian cluster of its own and is therefore immobile.   This observation is consistent with the expectation that locally isolated fractons are immobile~\cite{NandkishoreHermeleFractonsannurev-conmatphys-031218-013604,PretkoChenYou_2020fracton}.  A distinct feature, however, is that even though the particles have vanishing velocities, their momenta $\{p_a(0)\}$ are  undetermined and can be arbitrarily fixed. The disassociation between energy, velocities, and momenta has two  origins. The first is the exact microscopic symmetry as discussed in \cref{sec:Mach} -- dipole conservation results in a many-to-one relationship between momenta and velocity. When the initial separation of particles is large, however, not only is the total dipole moment $D = \sum_a x_a$ conserved, the coordinate of each particle $x_a$ is itself conserved and the conserved total dipole moment \emph{fragments} to $N$ conserved positions, $x_a$. 
 
 The second reason for the velocity-momentum-energy mismatch is emergent symmetries. The conserved quantities $x_a$ now generate a larger symmetry wherein each momentum can be arbitrarily shifted $p_a \mapsto p_a + \phi_a$. This property holds as we move away from the $U(x)\rightarrow 0,\eta \rightarrow \infty$ limit. When particles are well separated compared to the length scales set by $K(x)$ and $U(x)$ and the momentum differences are such that 
	\[\left(p_{a} - p_b\right)^2 K(x_a - x_b) \rightarrow 0,\] the Hamiltonian \cref{eq:Hamiltonian_classical_1d} vanishes and the discussion above holds.  
	
	For particles starting with a close initial separation, the situation is more complex. By numerically solving Hamilton's equations of motion for the Hamiltonian in \cref{eq:Hamiltonian_classical_1d}, we will see that velocity-momentum mismatch as well as emergent conserved quantities will generically appear accompanying qualitatively new kinds of trajectories. 
 
    Before we proceed, we point out  a useful canonical transformation  $\{x_j,p_j\} \mapsto \{q_\alpha,\pi_\alpha\}$:
	\begin{align}
		q_{\alpha = 1,\ldots,N-1} &= \frac{\sum_{j=1}^\alpha x_j - \alpha x_{\alpha + 1}}{\sqrt{\alpha \left(\alpha + 1\right)}}, ~q_N = \frac{\sum_{j=1}^N x_j}{\sqrt{N}}, \nonumber\\
		\pi_{\alpha = 1,\ldots,N-1} &= \frac{\sum_{j=1}^\alpha p_j - \alpha p_{\alpha + 1}}{\sqrt{\alpha \left(\alpha + 1\right)}}, ~\pi_N = \frac{\sum_{j=1}^N p_j}{\sqrt{N}}. \label{eq:reduced_coordinates}
	\end{align}
	Notice that $q_N = D/\sqrt{N}$ and $\pi_N = P/\sqrt{N}$ are  conserved quantities. These drop out when we express the Hamiltonian (\ref{eq:Hamiltonian_classical_1d})  using \cref{eq:reduced_coordinates}  making the translation and dipole symmetries manifest,
	\begin{multline}
		H(x_1, \ldots x_N ;p_1,\ldots p_N) = H(q_1,\ldots q_{N-1};\pi_1,\ldots \pi_{N-1})\\ \implies \{H,q_N\} = \{H,\pi_N\} = 0.
	\end{multline}
	These new coordinates help us reduce the effective phase space of $N$ particles to that of $N-1$ particles. We will refer to $\{q_\alpha, \pi_\alpha\}$ as the reduced phase space coordinates.
	
	\subsection{Two fractons on a line}
 \label{sec:2particles}
	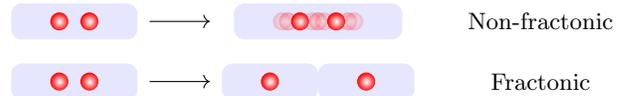
\begin{figure}[!ht]
		\begin{tikzpicture}[scale=0.8]
			
	{\fill[rounded corners,blue,opacity=0.1] (-0.8,-0.3+1) rectangle (1.3,0.3+1);}
	\foreach \x in {0,.5} 
	{	\shade[inner color=white, outer color = red] (\x ,1) circle (0.15);}
	
	\draw[->] (1.5,1) -- (2.5,1);

\fill[rounded corners,blue,opacity=0.1] (2.9,-0.3+1) rectangle (5.7,0.3+1);
		
			\foreach \x in {3.8,4.3,4.8} 
	{	\shade[inner color=white, outer color = red,opacity=0.3] (\x ,1) circle (0.15);}		

			\foreach \x in {3.7,4.2,4.4,4.9} 
{	\shade[inner color=white, outer color = red,opacity=0.2] (\x ,1) circle (0.15);}		

\foreach \x in {4.0,4.6} 
{	\shade[inner color=white, outer color = red] (\x ,1) circle (0.15);}

			{\fill[rounded corners,blue,opacity=0.1] (-0.8,-0.3) rectangle (0.5+0.8,0.3);}
			\foreach \x in {0,.5} 
			{	\shade[inner color=white, outer color = red] (\x ,0) circle (0.15);}
			\draw[->] (1.5,0) -- (2.5,0);
			\foreach \x in {4-0.5,5.6-0.5} 
			{\fill[rounded corners,blue,opacity=0.1] (\x-0.8,-0.3) rectangle (\x+0.8,0.3);}
			\foreach \x in {4-0.5,5.6-0.5} 
			{	\shade[inner color=white, outer color = red] (\x ,0) circle (0.15);}
   		\node[] at (8,0)  {Fractonic};				
     \node[] at (8,1)  {Non-fractonic};			
		\end{tikzpicture}
		\caption{Two closely separated particles in the same Machian cluster  exhibit two classes of trajectories. For  negative energies which are possible when interactions are attractive ($U<0$), the particles exhibit regular trajectories where they oscillate out of phase preserving the center of mass. For positive energies, they exhibit novel fractonic trajectories where they eventually separate and freeze to form two adjacent clusters. \label{fig:1d_2particles}}
	\end{figure}
	\subsubsection{Trajectories}
	
	\begin{figure}[!ht]
		\centering
		\begin{tabular}{c} 
			\includegraphics[width=0.35\textwidth]{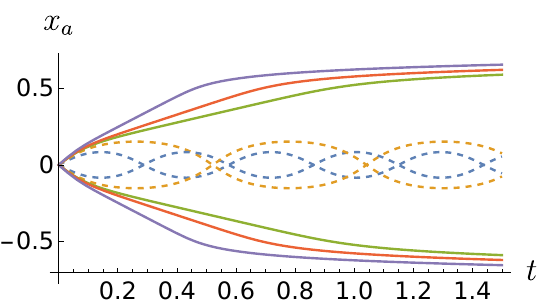} \\
			\includegraphics[width=0.35\textwidth]{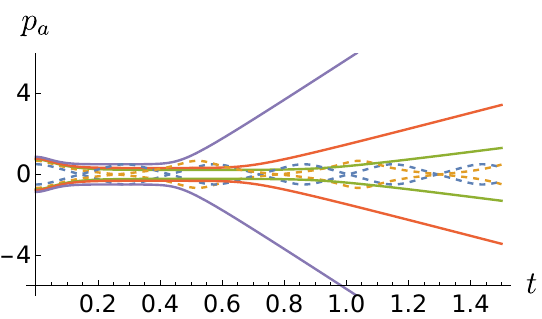} 	
		\end{tabular}
		\caption{Two particle trajectories generated by \cref{eq:Hamiltonian_classical_1d_2particles} for attractive interactions ($a=1,b=0.2,\eta = 10,\Gamma = -1$ ).  Negative energies (dotted, $E = -0.5, -0.1$) lead to oscillatory trajectories with bounded position and momentum coordinates.  Positive energies (solid, $E=0.1, 0.2, 0.5$) lead to `fractonic' trajectories where the position coordinates freeze out while the momenta grow indefinitely. }
		\label{fig:classical_2particles}
	\end{figure}		
	
	We consider two particles within a single Machian cluster i.e. such that $K(x_1(0) - x_2(0))\neq 0$. To analyse the ensuing dynamics, we begin by expressing its Hamiltonian in the reduced coordinates (\ref{eq:reduced_coordinates}) to eliminate the conserved dipole moment and total momentum, $q_2,\pi_2$ thereby reducing the phase space to that of a single degree of freedom with   Hamiltonian: 	
        \begin{equation}
		H = \pi_1^2 K(\sqrt{2} q_1) + U(\sqrt{2} q_1)  \label{eq:Hamiltonian_classical_1d_2particles}.
	\end{equation}
 and equations of motion
 \begin{align}
     \dot{q}_1 &= 2 \pi_1 K(\sqrt{2} q_1) \nonumber,\\\dot{\pi}_1 &= - \sqrt{2} \left( \pi_1^2 K'(\sqrt{2} q_1) + U'(\sqrt{2} q_1) \right). \label{eq:1deom}
 \end{align}

This system has a conserved quantity, the energy
$$E = \pi_1^2 K(\sqrt{2} q_1) + U(\sqrt{2} q_1)$$, which can be used to solve the equations of motion by quadrature:
	\begin{align}
		&t =  \int_{q(0)}^{q(t)} \frac{dy}{2\sqrt{K(\sqrt{2} y) \left(E - U(\sqrt{2} y)\right)}}, \nonumber \\
		&	\pi(t) =  \sqrt{\frac{E - U(\sqrt{2} q(t))}{K(\sqrt{2} q(t))}}. \label{eq:2_quadrature}
	\end{align}

	There are two possible classes of trajectories depending on the energy of the system and the nature of interactions. First, For  negative energy configurations, available with attractive interactions   ($\Gamma < E < 0$), we obtain familiar oscillating solutions, shown in  \cref{fig:classical_2particles} with dotted lines. The two particles stay within a single Machian cluster and oscillate perfectly out of phase to keep the center of mass fixed. Both position and momentum coordinates are bounded. For small oscillations, the frequency $\omega_0$ can be determined by setting $K \approx 1$ and Taylor expanding $U$ in \cref{eq:Hamiltonian_classical_1d_2particles} to obtain
	\begin{equation}
		H \approx \pi_1^2 + \frac{2 |\Gamma|}{b^2} x^2 \implies \omega_0 \approx \sqrt{\frac{8 |\Gamma|}{b^2}}.
	\end{equation}
 
Second,	for positive energy configurations, we find a novel class of trajectories. Irrespective of the nature of interactions (attractive $U<0$, repulsive $U>0$ or  no interactions $U=0$),  all trajectories with $E>0$ follow a similar pattern:  particles starting within a single Machian cluster eventually drift apart in opposite directions to keep the center of mass fixed, and eventually become immobile by forming two adjacent clusters. One curious aspect of these trajectories is that as the particles separate we have $K(x_1 - x_2) \rightarrow 0$ and $U(x_1 - x_2) \rightarrow 0$ in (\ref{eq:Hamiltonian_classical_1d_2particles}). The system conserves energy by $|p_1-p_2| \rightarrow \infty$ at late times appropriately. We call these trajectories `fractonic' to distinguish from ordinary trajectories present in dipole non-conserving systems where diverging momenta are not typically found. Sample fractonic trajectories are shown  \cref{fig:classical_2particles} with solid lines.

	\subsubsection{Dissipative dynamics and attractors}	
	\begin{figure}[!ht]
		\centering
			\begin{tabular}{c} 
			\includegraphics[width=0.35\textwidth]{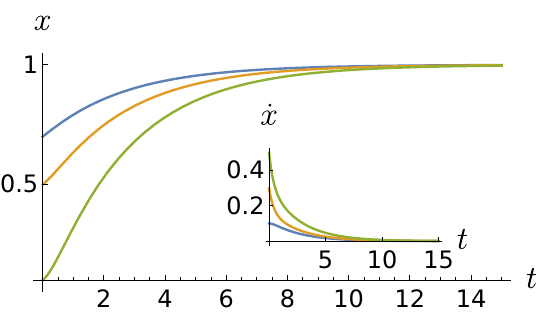} \\
                \includegraphics[width=0.35\textwidth]{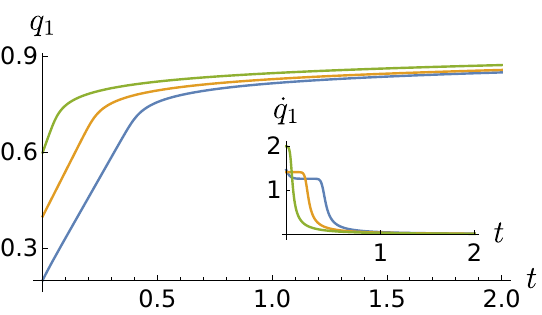}    
				\end{tabular}
		\caption{$\{ x, \dot{x}\}$ trajectories (top) for the over-damped oscillator governed by \cref{eq:2_dissipative}  with $m=k=a=1,\mu=3$ compared with the  fractonic trajectories (bottom) corresponding to positive energies $E = 0.4, 0.5,1.0$  for the two-particle system generated by the Hamiltonian in \cref{eq:Hamiltonian_classical_1d_2particles} with $a=1, b=0.2, \eta = 20$ and $\Gamma = -1$ in reduced (configuration) space coordinates $\{ q_1, \dot{q_1}\}$ .  }
		\label{fig:2_dissipation}
	\end{figure}	
It is instructive to compare the fractonic dynamics with that of a dissipative system. To do so, consider a simple one-dimensional classical damped oscillator given by 
	\begin{equation}
		m\ddot{x} =  - \mu \dot{x} -k (x-a), \label{eq:2_dissipative}
	\end{equation}
	whose trajectories  are shown in the top row of \cref{fig:2_dissipation} in the over-damped limit, $\mu > 2 \sqrt{k m}$ and for various initial initial conditions. We observe the well-known behaviour: the particle slows down in time and eventually becomes immobile. At a superficial level, this dynamics is qualitatively the same as that of two-particles discussed in \cref{sec:2particles} whose trajectories in the reduced coordinates are shown in the bottom row of \cref{fig:2_dissipation} for comparison. The crucial difference is that in the dissipative case, momentum is related to velocity in the usual way-- $p = m\dot{x}$ whereas in the case of fractons, it is not.
	\begin{figure}[!ht]
		\centering
			\begin{tabular}{c} 
			\includegraphics[width=0.35\textwidth]{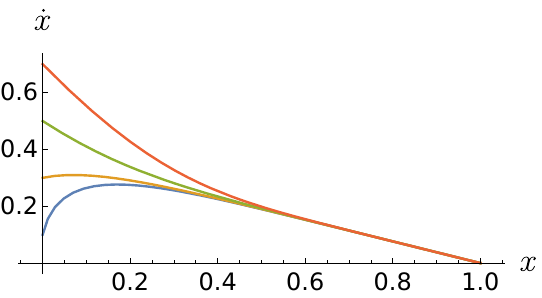} \\
			\includegraphics[width=0.35\textwidth]{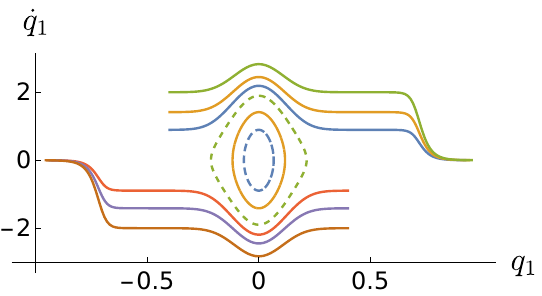} 
				\end{tabular}
		\caption{ Configuration space  portrait for the overdamped oscillator (top) with $m=k=a=1,\mu=3$ compared with the dipole conserving two-particle system in reduced coordinates (bottom) with $a=1, b=0.2, \eta = 20$ and $\Gamma = -1$. Both exhibit attractive fixed-points where multiple trajectories terminate. \label{fig:2_dissipation_attrctor}}
		
	\end{figure}
	
	 Dissipative dynamics is characterized by the presence of attractors in configuration (position, velocity) space: as $t \rightarrow \infty$, all trajectories approach the same global attractor $(x,\dot{x}) = (a,0)$ as shown in  \cref{fig:2_dissipation_attrctor}. However, for our dipole-conserving systems shown in \cref{fig:2_dissipation_attrctor}, the two classes of trajectories described above correspond to two types of fixed points. The first one at the origin $(q_1,\dot{q}_1) = \left(0 , 0 \right)$ is a \emph{center}~\cite{Strogatz2018nonlinear}  which is surrounded by  negative energy oscillatory trajectories. The second type of trajectories are, remarkably,  attractors located at $(q_1,\dot{q}_1) \approx \left(\pm \frac{a}{\sqrt{2}} , 0 \right)$ where all fractonic trajectories terminate. These trajectories are surprising because Liouville's theorem forbids such attractors in (position-momentum) phase space for Hamiltonian systems. Indeed, for ordinary (dipole non-conserving) Hamiltonian systems with $p = m \dot{x}$, Liouville's theorem implies that there are  no attractors in (position-velocity) configuration space. However, for fractonic trajectories in  dipole-conserving systems, we observe that the absence of attractors in phase space does not translate into the absence of attractors in configuration space. The key here is to realise that these attractors in configuration space are distinct trajectories in phase space.

\subsection{Three fractons on a line}
 
 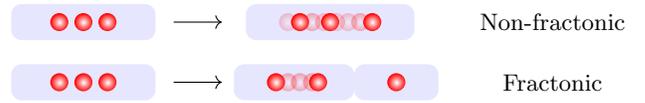
\begin{figure}[!ht]
		\begin{tikzpicture}[scale=0.8]
	{\fill[rounded corners,blue,opacity=0.1] (-1,-0.3+1) rectangle (1.4,0.3+1);}
	\foreach \x in {-0.2,0.2,.6} 
	{	\shade[inner color=white, outer color = red] (\x ,1) circle (0.15);}
	
	\draw[->] (1.7,1) -- (2.5,1);

\fill[rounded corners,blue,opacity=0.1] (2.9,-0.3+1) rectangle (5.7,0.3+1);
		
			\foreach \x in {4.0,4.8} 
	{	\shade[inner color=white, outer color = red,opacity=0.3] (\x ,1) circle (0.15);}		

			\foreach \x in {3.6,4.6,4.2,4.4} 
{	\shade[inner color=white, outer color = red,opacity=0.2] (\x ,1) circle (0.15);}		

\foreach \x in {3.8,4.3,5} 
{	\shade[inner color=white, outer color = red] (\x ,1) circle (0.15);}

			{\fill[rounded corners,blue,opacity=0.1] (-1,-0.3) rectangle (1.4,0.3);}
			\foreach \x in {-0.2,0.2,.6}
			{	\shade[inner color=white, outer color = red] (\x ,0) circle (0.15);}
			\draw[->] (1.7,0) -- (2.5,0);
			
			{\fill[rounded corners,blue,opacity=0.1] (3.5-0.8,-0.3) rectangle (3.9+0.8,0.3);}

   {\fill[rounded corners,blue,opacity=0.1] (5.5-0.8,-0.3) rectangle (5.3+0.8,0.3);}
   			\foreach \x in {3.6,3.8,4.0} 
			{	\shade[inner color=white, outer color = red,opacity=0.3] (\x ,0) circle (0.15);}
			\foreach \x in {3.4,4.1,5.4} 
			{	\shade[inner color=white, outer color = red] (\x ,0) circle (0.15);}
   		\node[] at (8,0)  {Fractonic};				
     \node[] at (8,1)  {Non-fractonic};			
		\end{tikzpicture}
		\caption{Three closely separated particles in the same Machian cluster  exhibit three classes of trajectories. For negative energies which are possible when interactions are attractive ($U<0$), the particles exhibit regular oscillations or chaos preserving the center of mass. For positive energies, they exhibit fractonic trajectories where two particles settle to an oscillatory state in one cluster while the third freezes out in another. \label{fig:1d_3particles}}
	\end{figure}
 
	\label{sec:3particles}

	\begin{figure}[!ht]
		\centering
		\begin{tabular}{c}
			\includegraphics[width=0.49\textwidth]{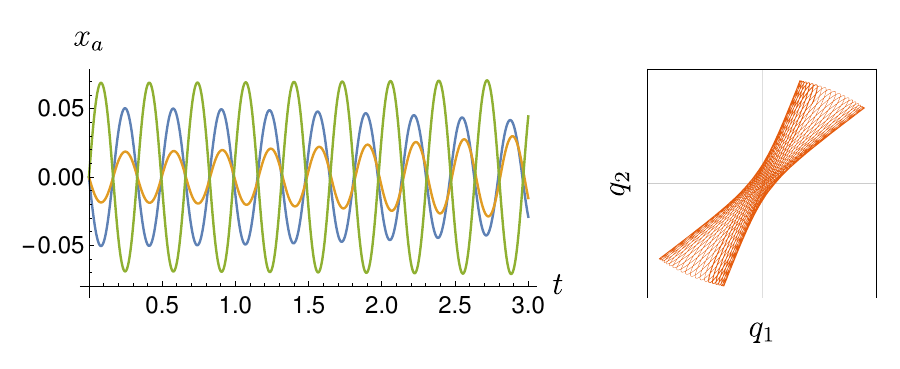} \\
   \includegraphics[width=0.49\textwidth]{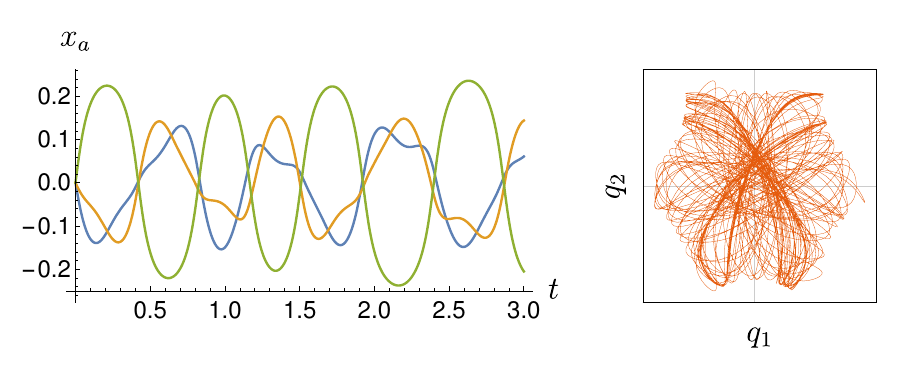} 
		\end{tabular}
		\caption{Non-fractonic regular (top, $E=-2.5$) and chaotic (bottom, $E=-1$) trajectories of three particle position ($x_a$) coordinates along with corresponding two-dimensional phase portraits in reduced coordinates ($q_1 - q_2$) generated by the Hamiltonian in \cref{eq:Hamiltonian_classical_1d_3particles} with $a=1, b=0.2, \eta=10$ and $\Gamma=-1$.}
		\label{fig:classical_3particles_regular}
	\end{figure}	
	
We now consider the case of three particles. The Hamiltonian (\ref{eq:Hamiltonian_classical_1d}) for $N=3$ in reduced coordinates \cref{eq:reduced_coordinates} contains two effective degrees of freedom leading to dynamics in the four dimensional phase-space $(q_{1},\pi_{1},q_2,\pi_2)$ with Hamiltonian: 
	\begin{multline}
		H =   \pi_1^2 K(\sqrt{2} q_1) +   \frac{\left(\sqrt{3}\pi_2 + \pi_1\right)^2}{4}K \left(\frac{\sqrt{3}q_2 + q_1}{\sqrt{2}}\right)  \\
		+ \frac{\left(\sqrt{3}\pi_2 - \pi_1\right)^2}{4} K \left(\frac{\sqrt{3}q_2 - q_1}{\sqrt{2}}\right)+  U(\sqrt{2} q_1) \\+  U \left(\frac{\sqrt{3}q_2 + q_1}{\sqrt{2}}\right) + U \left(\frac{\sqrt{3}q_2 - q_1}{\sqrt{2}}\right). \label{eq:Hamiltonian_classical_1d_3particles}
	\end{multline}

\subsubsection{Trajectories}
Just like in the case of two particles we observe again non-fractonic trajectories for certain negative energies with attractive interactions ($\Gamma<0$). However, unlike the two-particle case, the dynamics can now be regular or chaotic as shown in \cref{fig:classical_3particles_regular}. The appearance of non-fractonic chaotic and regular trajectories in a system with two degrees of freedom is familiar and qualitatively the same as the famous H\'{e}non-Heiles model~\cite{HenonHeiles1964applicability}. However, the latter has free trajectories where the particles travel with constant momenta whereas out system  supports instead fractonic trajectories.

	\begin{figure}[!ht]
		\centering
		\begin{tabular}{c}
			\includegraphics[width=0.5\textwidth]{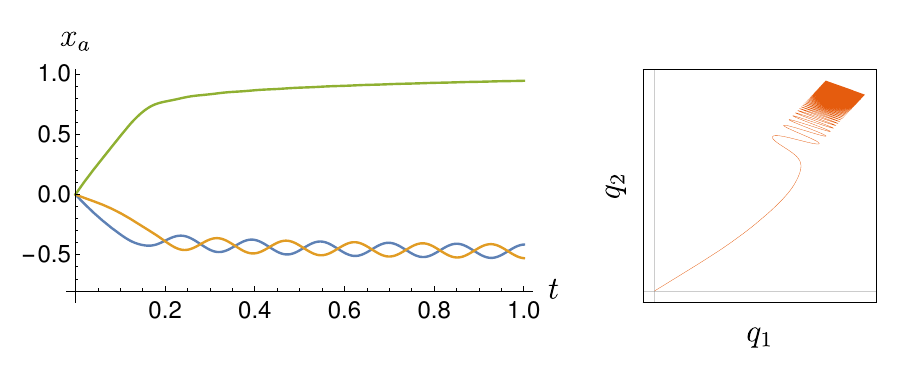} \\
			\includegraphics[width=0.5\textwidth]{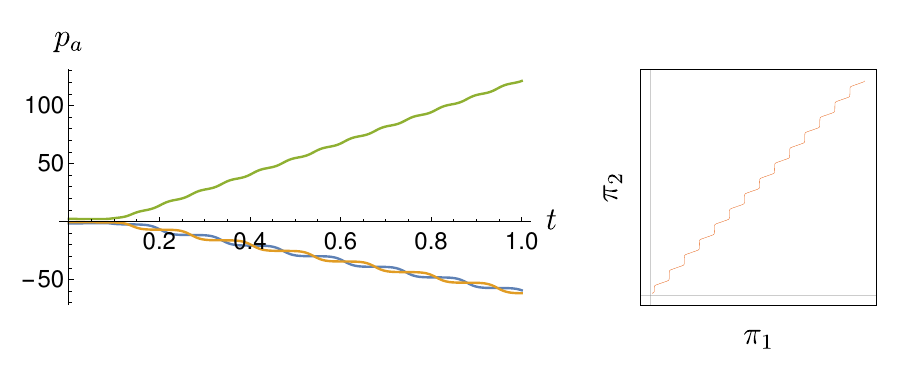} 
		\end{tabular}
		\caption{ Fractonic trajectories ($E=10$) of three particle position ($x_a$) and momentum ($p_a$) coordinates along with corresponding two-dimensional phase portraits in reduced coordinates ($q_1 , q_2$) generated by the Hamiltonian in \cref{eq:Hamiltonian_classical_1d_3particles} with $a=1, b=0.2, \eta=10$ and $\Gamma=-1$.}
		\label{fig:classical_3particles_fracton}
	\end{figure}

     Indeed, fractonic trajectories are found for both positive and negative energy configurations and therefore for attractive ($\Gamma<0$), repulsive ($\Gamma >0$) or indeed no interactions ($\Gamma = 0$). These are qualitatively different from the fractonic trajectories of the two-particle case. A direct extrapolation of the two-particle picture suggests that at late times, we expect the particles starting within each other's presence in a single cluster to separate out until they are isolated within their own separate Machian clusters and become immobile. In fact, this scenario only occurs for fine-tuned initial conditions as will be proven within an exactly solvable limit in \cref{sec:exact_trajectories}. As shown in \cref{fig:classical_3particles_fracton}, we see that generically, fracton trajectories corresponds to the system breaking up with one of the particles separating out from the other two and eventually becoming immobile. In this scenario, the remaining two particles form a single cluster and settle down to an indefinite oscillatory state about their common center of mass. Interestingly, these oscillations result from the particles bouncing off the edge of the Machian cluster of the first particle. Hence this two-body dynamics requires interaction with the third particle.

       In summary, generically, when three particles start off within a single cluster, fractonic trajectories lead to the formation of at least two separate clusters. For certain fine-tuned conditions as will be discussed in \cref{sec:exact_trajectories}, fractonic trajectories lead to three isolated clusters.

\subsubsection{Limit cycles}
 \begin{figure}[!ht]
		\centering
		\begin{tabular}{c}
			\includegraphics[width=0.49\textwidth]{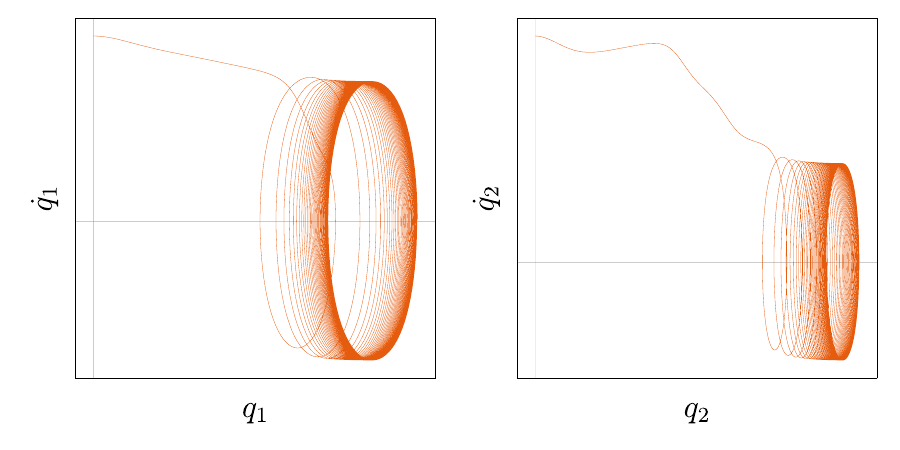} 
		\end{tabular}
		\caption{Projections of the reduced coordinate configuration phase portrait of the fractonic trajectory shown in \cref{fig:classical_3particles_fracton} exhibits a limit-cycle behaviour.}
		\label{fig:classical_3limit}
	\end{figure}	

	We saw in \cref{sec:2particles} that fractonic trajectories for two particles ended up in an attractor in the position-velocity configuration space. For three particles, generic fracton trajectories asymptotically end up in a limit cycle in configuration space as shown in \cref{fig:classical_3limit}. Just like attractors, Liouville's theorem forbids the presence of limit cycles in phase space which, for non-dipole-conserving systems where velocities are proportional to momenta, also implies the absence of limit cycles in configuration space. For fractons, this is no longer true and while \emph{mathematical} limit cycles are not to be found in phase space, consistent with Liouville's theorem, \emph{physical} limit cycles do appear in configuration space! 
 
	\subsection{Four and more fractons on a line}
	\label{sec:4particles}
	\begin{figure}[!ht]
		\begin{tikzpicture}[scale=0.8]
	{\fill[rounded corners,blue,opacity=0.1] (-1-0,-0.3-1) rectangle (1.8,0.3-1);}
	\foreach \x in {-0.2,0.2,.6,1} 
	{	\shade[inner color=white, outer color = red] (\x ,-1) circle (0.15);}
	
	\draw[->] (2,-1) -- (2.8,-1);

\fill[rounded corners,blue,opacity=0.1] (3,-0.3-1) rectangle (5.6,0.3-1);
		{\fill[rounded corners,blue,opacity=0.1] (5.5-0.8+0.9,-0.3-1) rectangle (5.3+0.6+0.9,0.3-1);}
			\foreach \x in {4.0,4.8} 
	        {	\shade[inner color=white, outer color = red,opacity=0.3] (\x ,-1) circle (0.15);}	
			\foreach \x in {3.6,4.6,4.2,4.4} 
                {	\shade[inner color=white, outer color = red,opacity=0.2] (\x ,-1) circle (0.15);}		
                \foreach \x in {3.8,4.3,5,6.2} 
                {	\shade[inner color=white, outer color = red] (\x ,-1) circle (0.15);}
   {\fill[rounded corners,blue,opacity=0.1] (-1,-0.3) rectangle (1.8,0.3);}
	\foreach \x in {-0.2,0.2,.6,1} 
	{	\shade[inner color=white, outer color = red] (\x ,0) circle (0.15);}
			\draw[->] (2,0) -- (2.8,0);
			
			{\fill[rounded corners,blue,opacity=0.1] (3.5-0.8 +0.3,-0.3) rectangle (3.9+0.8+0.3,0.3);}
   			\foreach \x in {3.6,3.8,4.0} 
			{	\shade[inner color=white, outer color = red,opacity=0.3] (\x +0.3,0) circle (0.15);}
			\foreach \x in {3.4,4.1} 
			{	\shade[inner color=white, outer color = red] (\x+0.3 ,0) circle (0.15);}

    			{\fill[rounded corners,blue,opacity=0.1] (3.5-0.8 +0.3 +2,-0.3) rectangle (3.9+0.8+0.3+2,0.3);}
   			\foreach \x in {3.6,3.8,4.0} 
			{	\shade[inner color=white, outer color = red,opacity=0.3] (\x +0.3+2,0) circle (0.15);}
			\foreach \x in {3.4,4.1} 
			{	\shade[inner color=white, outer color = red] (\x+0.3+2 ,0) circle (0.15);}
       
		\end{tikzpicture}
		\caption{Four closely separated particles in the same Machian cluster  exhibit three classes of fractonic trajectories. Asymptotically, the system can separate into two clusters with  two particles each  where each pair exhibits oscillatory trajectories (top) or one frozen particle in one cluster and three in the other, exhibiting either regular oscillations or chaos  (bottom).\label{fig:1d_4particles}}
	\end{figure}
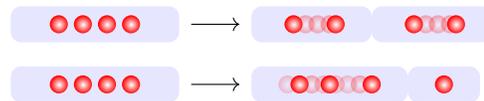

	We now turn to the case of four particles. The Hamiltonian written in reduced coordinates is now a function of six phase space coordinates corresponding to three degrees of freedom. We will not write the explicit form here. With attractive interactions, we again obtain non-fractonic bounded oscillatory and chaotic trajectories qualitatively similar to the three-particle case shown in \cref{fig:classical_3particles_regular}. We will therefore switch off the interactions ($U(x) = 0$) for the rest of this section and focus only on fractonic trajectories. 

 	\begin{figure}[!ht]
		\centering
		\begin{tabular}{cc}
			\includegraphics[width=0.239\textwidth]{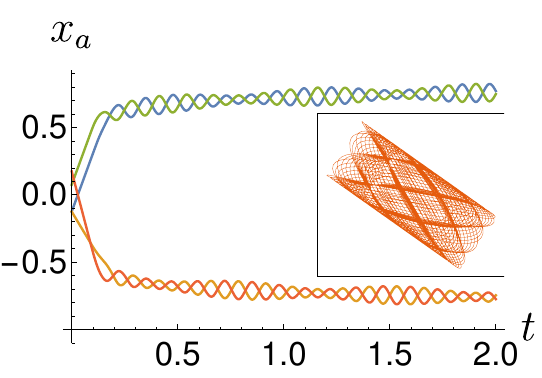}& 
			\includegraphics[width=0.239\textwidth]{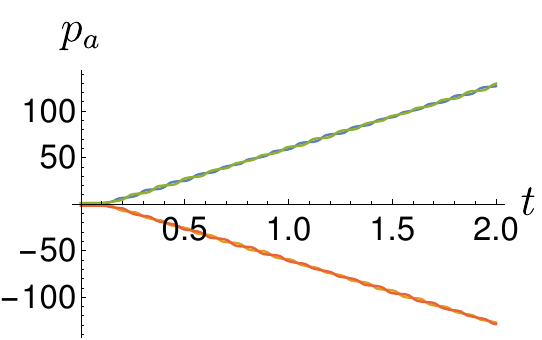}\\ 
   			\includegraphics[width=0.239\textwidth]{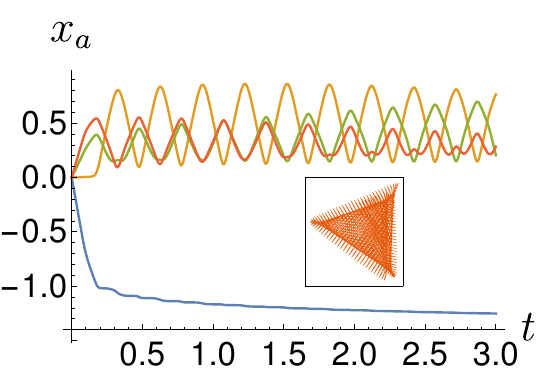}& 
			\includegraphics[width=0.239\textwidth]{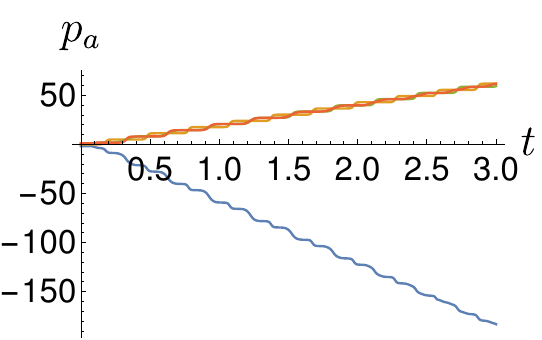}\\ 
			\includegraphics[width=0.239\textwidth]{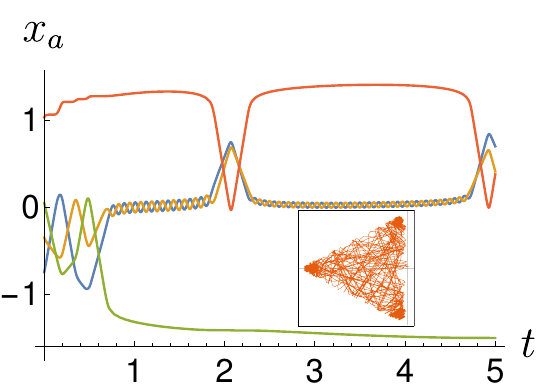} &
			\includegraphics[width=0.239\textwidth]{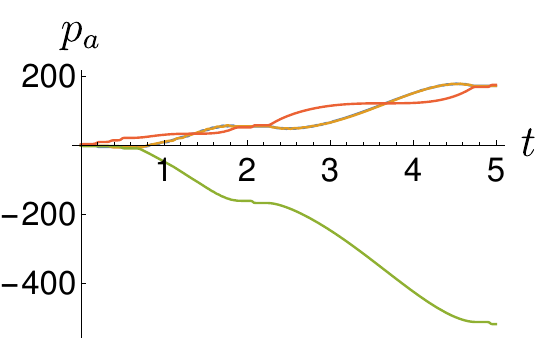} 
		\end{tabular}
		\caption{Fractonic regular trajectories exhibiting three-particle oscillations (top), two particle oscillations (middle) and chaos (bottom) generated by the four-particle Hamiltonian without interactions \cref{eq:Hamiltonian_classical_1d} with $E=10$ and $a = 1, \eta = 10$ and  $\Gamma = 0$. The insets show reduced coordinate $q_1,q_2$ projections of phase portraits where the regular and chaotic nature of trajectories are clearer.}
		\label{fig:classical_4particles_fracton}
	\end{figure}	

    As shown in \cref{fig:classical_4particles_fracton}, we find that four particles starting within a single Machian cluster eventually separate into two. The nature of asymptotic dynamics in each of the two asymptotic clusters now depends sensitively on initial conditions.  \Cref{fig:classical_4particles_fracton},  shows samples of the various trajectories possible for the same fixed energy and Hamiltonian parameters but different initial conditions. We see two classes of regular fractonic trajectories (top two rows) where the four particles split into subsystems with two-two or three-one particles and the multi-particle subsystems settle down into indefinite oscillations at late times. These are qualitatively similar to the fractonic trajectories for three particles shown in \cref{fig:classical_3particles_fracton}. The bottom panel in \cref{fig:classical_4particles_fracton} shows a new class of fracton trajectories where the four particles split into three-one particle subsystems and the three particles exhibit chaos. Like the regular fracton trajectories, the chaotic fracton trajectory has momenta that grow indefinitely with time. 

 We can conjecture that the trajectories studied above for four or fewer particles qualitatively exhausts all possibilities for a finite number of fractons -- particles within a single Machian cluster exhibit either (i) non-fractonic regular or chaotic dynamics in the same cluster or (ii) fractonic regular or chaotic dynamics where the system asymptotically splits into multiple Machian clusters. We have checked this conjecture numerically for systems with multiple number of particles starting in close proximity and several randomly generated initial conditions. 
 
	\section{Oppositely charged fractons on a line}
	\label{sec:Classical trajectories Unlike charges}	
        We now relax the conditions of \cref{sec:Classical trajectories} and consider the case where not all particles carry the same charge. We will only consider the case with no interactions i.e. $U(x)=0$. The Hamiltonian in \cref{eq:Hamiltonian_classical_1d} is modified to 
        \begin{equation}
		H = \frac{1}{2} \sum_{a<b=1}^N  \left( \frac{p_a}{q_a} - \frac{p_b}{q_b} \right)^2 K(x_{a} - x_b). \label{eq:Hamiltonian_classical_1d_unlike}
	\end{equation}
        As discussed in \cref{sec:Symmetries_and_compatible_hamiltonians}, the conserved dipole moment and the symmetry  transformation induced by it are
        \begin{equation}
            D = \sum_a q_a x_a,~\dot{D} = 0 \implies p_a \mapsto p_a + q_a \phi.
        \end{equation}
        For concreteness, we consider the charge assignment $q_a = (-1)^a$ so that we have two species of particles with charges $\pm 1$ and the form of $K(x)$ as shown in \cref{eq:K(x) regulated}. The whole system is net charge neutral for even number of particles. The main new possibility that opens up is that of particles pairing up into charge neutral dipoles which can in principle behave like ordinary non-fractonic particles. The simplest case to consider is two particles.
  \subsection{Two oppositely charged fractons}
  \label{sec:2particles_unlike}
        The two-particle version of the Hamiltonian (\ref{eq:Hamiltonian_classical_1d_unlike}) is
	\begin{equation}
		H = \frac{1}{2} \left( p_1 + p_2 \right)^2 K(x_1 - x_2).
	\end{equation}
	This Hamiltonian is entirely a function of the conserved quantities-- the total dipole moment $D = x_1 - x_2$ and  momentum $P = p_1 + p_2$
	\begin{equation}
		H = \frac{P^2}{2} K(D).
	\end{equation}
	The equations of motion can be trivially solved to obtain
	\begin{align}
		x_a(t) &= x_a(0) + P K(D) t, \nonumber \\ 
		p_a(t) &= p_a(0) - (-1)^a \frac{P^2}{2} K'(D) t. \label{eq:eom_unlike_general}
	\end{align}
	The two particles moves together as a dipole unit with a fixed separation and with a constant velocity, $v_{\text{eff}} = P K(D)$. In other words, the dipole moves as a free particle with an effective dipole moment- dependent mass $m_{\text{eff}} = K(D)^{-1}$. For larger separation between the two particles,   the dipole moment increase and so does the mass. Particles with a large separation therefore   become  immobile. This qualitative picture for the motion of opposite-charged pair of fractons was also observed in \cite{Pretko_MachPhysRevD.96.024051}. Finally, we note that while the momentum of the center of mass is constant, individual momenta evolve with time as shown in \cref{eq:eom_unlike_general} and increase indefinitely with time. 

    \subsection{Many oppositely charged fractons}
    	\begin{figure}[!ht]
		\centering
		\begin{tabular}{lr}
			\includegraphics[width=0.23\textwidth]{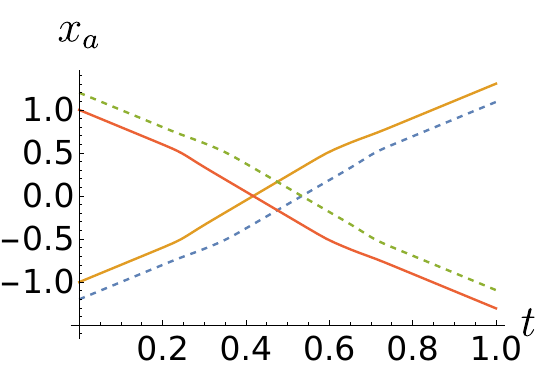} &
			\includegraphics[width=0.23\textwidth]{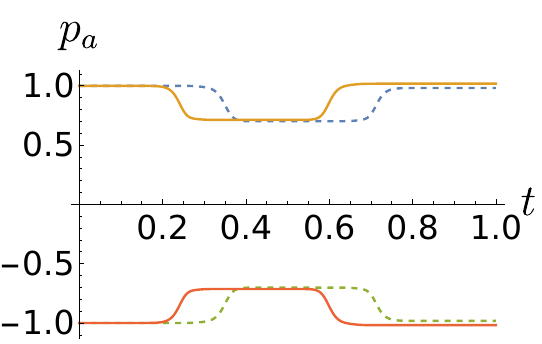} \\
   			\includegraphics[width=0.23\textwidth]{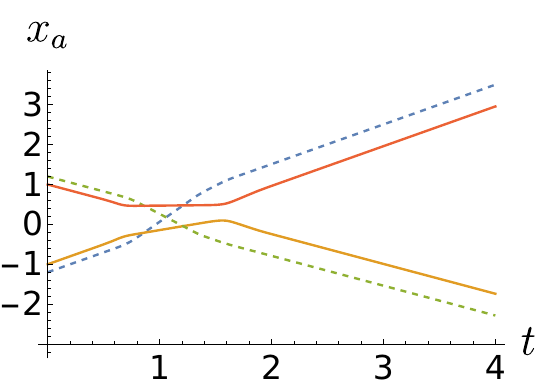} &
			\includegraphics[width=0.23\textwidth]{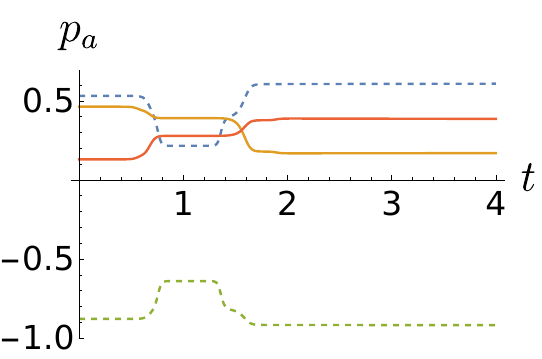} 
		\end{tabular}
		\caption{ Trajectories generated by the four particle Hamiltonian in \cref{eq:Hamiltonian_classical_1d_unlike} with $a=1$ and $\eta=10$ showing asymptotic free dipole motion. Positive (negative) charge motion is described with solid (broken) lines. When the dipoles meet, they may not (top)  or may (bottom) exchange partners. }
		\label{fig:4particles_fracton_unlike}
	\end{figure}
\begin{figure}[!ht]
		\centering
		\begin{tabular}{lr}
   		\includegraphics[width=0.23\textwidth]{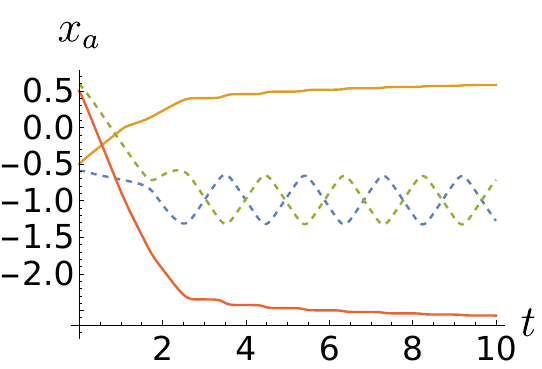} &
			\includegraphics[width=0.23\textwidth]{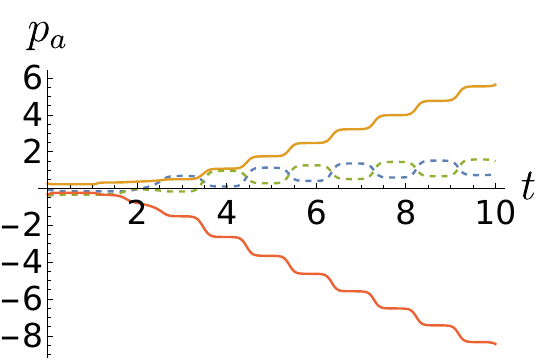} \\
   \includegraphics[width=0.23\textwidth]{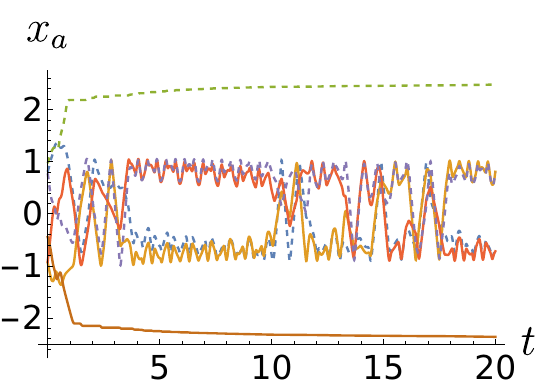} &
			\includegraphics[width=0.23\textwidth]{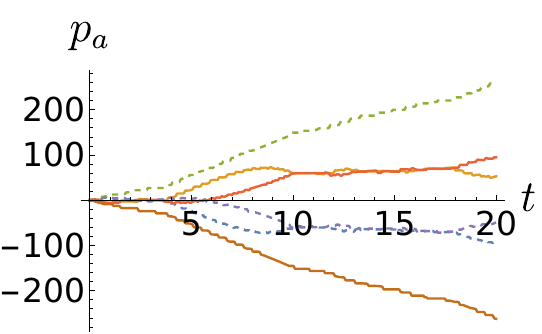} \\
   \includegraphics[width=0.23\textwidth]{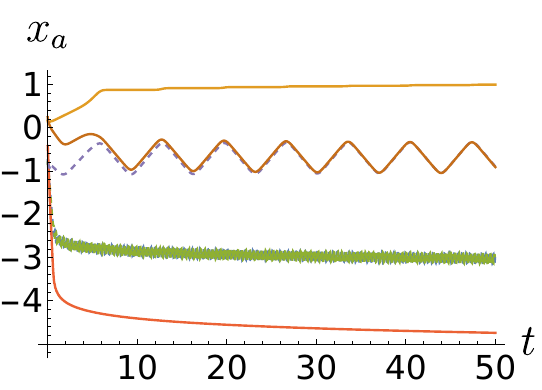} &
			\includegraphics[width=0.23\textwidth]{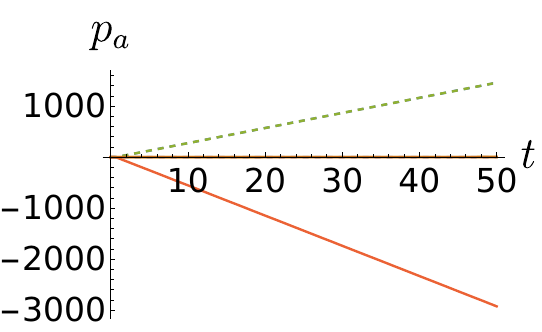} \\
		\end{tabular}
		\caption{ Fractonic trajectories generated by the four particle (top)  and six particle (middle, bottom)  Hamiltonians of \cref{eq:Hamiltonian_classical_1d_unlike} with $a=1$ and $\eta=10$. Positive (negative) charge motion is described with solid (broken) lines. We see the formation of multiple clusters each of which exhibits fractonic regular and chaotic motion with unbounded momentum growth.}
		\label{fig:6particles_fracton_unlike}
	\end{figure}

    With a larger number of particles, new possibilities emerge. One straightforward generalization of the two-particle case is that dipoles move as free particles in isolation from other particles. However, when two dipoles meet, they can interact through the locality term $K(x)$. They can either (i) pass through each other when they meet as shown in the top column of \cref{fig:4particles_fracton_unlike} or (ii) exchange partners by swapping particles of like charge between dipole units as shown in the bottom column of \cref{fig:4particles_fracton_unlike}. However, more complex trajectories are also possible wherein the system clusters in various ways and like particles exhibit fractonic trajectories described in \cref{sec:Classical trajectories}. Samples of these complex trajectories are shown in \cref{fig:6particles_fracton_unlike}.
	\section{Exact solutions}
	\label{sec:exact_trajectories}	
	In this section, we consider a limit wherein fracton trajectories of like particles can be obtained exactly. This will confirm the numerical trajectories shown in \cref{sec:Classical trajectories}. For this, we consider the Hamiltonian (\ref{eq:Hamiltonian_classical_1d}), switch off interactions $U(x) =0$, and set $\eta \rightarrow \infty$ to obtain
	\begin{equation}
		H = \frac{1}{2}  \sum_{a<b}B(x_a - x_b)  \left(p_a- p_b\right)^2  \label{eq:Hamiltonian_exactlysolvable}
	\end{equation} 	
	where, $B(x)$ is the as defined in \cref{eq:K(x) box}. For this Hamiltoinain, we can obtain exact, regular fractonic trajectories for up to three particles. Since there exists a solution by quadrature given in \cref{eq:2_quadrature} for the two-particle Hamiltonian (\ref{eq:Hamiltonian_classical_1d_2particles}), the first non-trivial case is the three-particle case. We will discuss the solution for two-particles as a warm up and then proceed to three particles. 
	
	\subsection{Warm-up: Two particles}
	\label{sec:exact_2particles}
	\begin{figure}[!ht]
		\centering
		\begin{tabular}{cc} 
			\includegraphics[width=0.239\textwidth]{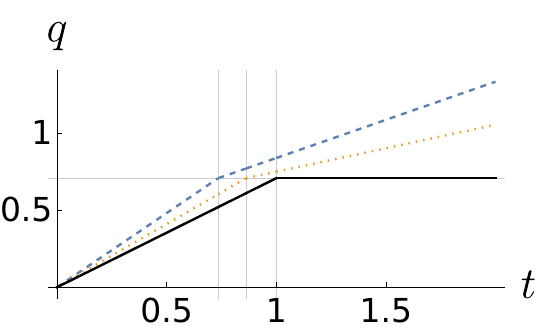}	&		\includegraphics[width=0.239\textwidth]{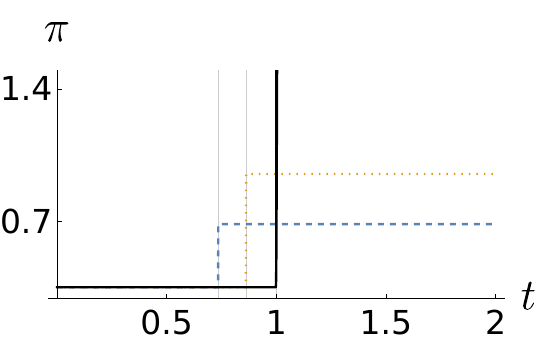}	
		\end{tabular}
		\caption{Plot of exact trajectories shown in \cref{eq:2_exact_regulated} for $\delta = 0.6$ (dashed line), $0.3$ (dotted line) and $0$ (solid line). Initial conditions chosen as $q(0) =0,~\pi(0)=(2 \sqrt{2})^{-1}$ and we set $a=1$\label{fig:Exact_2}. The gridlines  mark $t^* = \frac{a}{2 \sqrt{2} \pi(0)(1+\delta^2)}$ on $t$ axis and $q =a/\sqrt{2} $ on the $q$ axis.}
	\end{figure}	
	The Hamiltonian  (\ref{eq:Hamiltonian_exactlysolvable}) for two particles written in terms of the reduced coordinates  (\ref{eq:reduced_coordinates}) is 
	\begin{equation}
		H = \pi_1^2 B(\sqrt{2} q_1). \label{eq:2_particle_exact}
	\end{equation} 
	We will henceforth drop the redundant subscript  $\{q_1,\pi_1\} \equiv \{q,\pi\}$ to reduce clutter. We will only consider initial conditions such that \cref{eq:2_particle_exact} does not vanish i.e. $|q(0)| < a/\sqrt{2}$ and $\pi(0) \neq 0$. We want to solve Hamilton's equations of motion,
	\begin{equation}
		\dot{q}= 2 \pi B(\sqrt{2} q),~\dot{\pi} = -\sqrt{2}\pi^2  B'(\sqrt{2} q). \label{eq:2_eom_exact}
	\end{equation}
To do this, we first regulate $B(x)$ so that it does not vanish for any value of $x$ 
	\begin{equation}
		B(x) \rightarrow B(x) + \delta^2
	\end{equation}
	and eventually take $\delta \rightarrow 0$. Notice that when $|q| \neq a/\sqrt{2}$,  \cref{eq:2_eom_exact}  reduces to equations for free particles. Therefore, our strategy is to solve \cref{eq:2_eom_exact} piecewise and then stitch up the solutions. For $|q_1(0)| < a/\sqrt{2}$, we have $B'(\sqrt{2} q) =0$ and $B(\sqrt{2} q) = 1+\delta^2$ and \cref{eq:2_eom_exact} can be integrated to obtain
	\begin{equation}
		\pi(t) = \pi(0), ~q(t) = q(0) + 2\pi(0) (1+\delta^2) t. \label{eq:2_exact_early}
	\end{equation}
	This solution holds for $t<t^*$ where $t^*$ is defined by the condition 
	\begin{equation}
		|q(t^*)| = a/\sqrt{2} \label{eq:tstar}.
	\end{equation}
	For $t>t^*$,  we have $ |q| > a/\sqrt{2}$ and therefore $B'(\sqrt{2} q) =0$ and $B(\sqrt{2} q) = \delta^2$. \Cref{eq:2_eom_exact} can again be solved to obtain
	\begin{equation}
		\pi(t) = \pi_>, ~q(t) = q(t^*) + 2\pi_> \delta^2 t, \label{eq:2_exact_late_tbd}
	\end{equation}
	where, $q(t^*) = \text{sgn}(\pi(0)) a/\sqrt{2}$. $\pi_>$ can be determined using energy conservation.
	\begin{multline}
	E(q(t<t^*),\pi(t<t^*)) = E(q(t>t^*),\pi(t>t^*))  \\ \implies  \pi_> = \pi(0) \frac{\sqrt{1+\delta^2}}{\delta}.\label{eq:2_exact_pijump}
	\end{multline}
	Substituting into \cref{eq:2_exact_late_tbd} we obtain the full piecewise defined solution, 
	\begin{align}
		q(t) &= \begin{cases}
			q(0) + 2\pi(0) t (1+\delta^2) &t<t^*\\
			q(t^*) + 2\pi(0)(t-t^*)\delta \sqrt{1+\delta^2} &t>t^*
		\end{cases}\nonumber \\
		\pi(t) &= \begin{cases}
			\pi(0) &t<t^*\\
			\pi(0) \frac{\sqrt{1+\delta^2}}{\delta} &t>t^*
		\end{cases} \label{eq:2_exact_regulated}.
	\end{align}
	We can now take $\delta\rightarrow 0$ to remove the regulator and  obtain
	\begin{align}
		q(t) &= \begin{cases}
			q(0) + 2\pi(0)  t &t<t^*\\
			q(t^*) &t>t^*.
		\end{cases} 
		\nonumber \\
		\pi(t) &=  \begin{cases}
			\pi(0)~~~~t<t^*\\
			\infty~~~~~~~t>t^*
		\end{cases}
		\label{eq:2_exact_final}		
	\end{align}
 
	This solution  confirms qualitatively the nature of the two-particle fracton trajectories shown in \cref{sec:2particles}. 
	
	\subsection{Symmetries and distinct sectors for three particles}

	\begin{figure}
		\centering
			\includegraphics[width=0.3\textwidth]{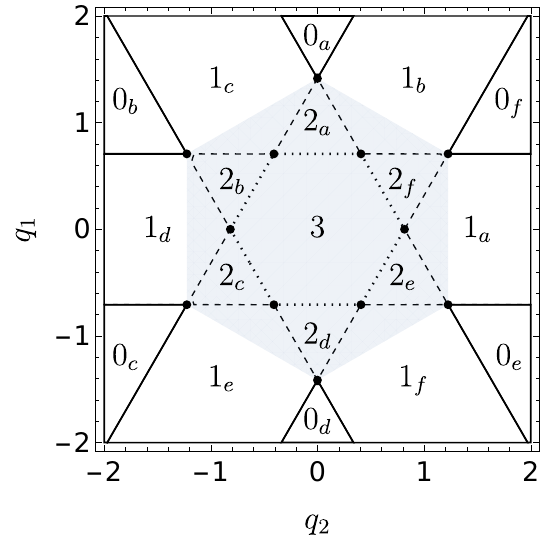}	
		\caption{ Distinct regimes of the three-particle Hamiltonian in \cref{eq:3_particles_exact}. The labels $M_a,M_b$ etc. indicate the number of particle pairs $M$ that are proximate such that \cref{eq:proximity} holds. The shaded region surrounding the origin represents the initial conditions that can lead to an oscillating steady state.  \label{fig:Regions_3_label}}
	\end{figure}

	Next, we consider the three-particle case. Then, the Hamiltonian (\ref{eq:Hamiltonian_exactlysolvable}) can be written as
 \begin{align}
     &H = H_{12} + H_{13} + H_{23},\\
 \nonumber    &H_{ab} \equiv \frac{1}{2} (p_a - p_b)^2 B(|x_a - x_b|).
 \end{align}
 In reduced coordinates, it corresponds to the the following Hamiltonian acting on the four-dimensional phase space of two degrees of freedom
	\begin{multline}	
		 H= \pi_1^2 B(\sqrt{2} q_1)+ \frac{\left(\sqrt{3}\pi_2 + \pi_1\right)^2}{4}B \left(\frac{\sqrt{3}q_2 + q_1}{\sqrt{2}}\right) \\ + \frac{\left(\sqrt{3}\pi_2 - \pi_1\right)^2}{4} B \left(\frac{\sqrt{3}q_2 - q_1}{\sqrt{2}}\right). \label{eq:3_particles_exact}
	\end{multline}
	  We will use the form in \cref{eq:3_particles_exact} for the rest of this section. We begin by dividing the two-dimensional position-space $\{q_1,q_2\}$ into different sectors depending on how many pairs of particles are proximate to each other i.e. the number of pairs $(a,b)$ such that
	\begin{equation}
		|x_a - x_b|<a  \implies B(|x_a - x_b| ) = 1. \label{eq:proximity}
	\end{equation}
 
This is summarized in \cref{fig:Regions_3_label} where the labels $0,1,2,3$ represent the number of pairs that satisfy \cref{eq:proximity}. The Hamiltonian  (\ref{eq:3_particles_exact}) has the symmetries of a regular hexagon  manifest in \cref{fig:Regions_3_label}, generated by reflections and six-fold rotations. The action of these symmetries on phase-space coordinates are
	\begin{align}
		& \{\vec{q} \mapsto R\vec{q},~\vec{\pi} \mapsto R \vec{\pi}\},~
		R= \begin{pmatrix}
			~~\cos \omega &   \sin \omega\\
			-\sin\omega &   \cos\omega  
		\end{pmatrix} \text{ and} \nonumber \\ &\{\vec{q} \mapsto X\vec{q},~\vec{\pi} \mapsto X\vec{\pi}\} ,~ X= \begin{pmatrix}
			1 & 0 \\
			0 & -1
		\end{pmatrix} , \label{eq:3_symmetry}
	\end{align}
	where $ \vec{q} \equiv \begin{pmatrix}
		q_1, q_2
	\end{pmatrix}^T,~\vec{\pi} \equiv \begin{pmatrix}
		\pi_1, \pi_2
	\end{pmatrix}^T$. 
 
 Together, $R$ and $X$ form an irreducible representation of the dihedral group $D_{12}$~\cite{Ramond2010group}.	The invariance of the Hamiltonian (\ref{eq:3_particles_exact}) under (\ref{eq:3_symmetry}) is useful because it maps the various sectors $\{M_{a} \ldots M_f\}$ shown in \cref{fig:Regions_3_label} into each other. Consequently, trajectories with one set of initial conditions can be used to generate trajectories with other initial conditions by symmetry transformations which allows us to restrict our attention to only a representative set of cases.  
 
 The Hamilton's equations of motion (eom) for \cref{eq:3_particles_exact} are
	\begin{align}
		\dot{q}_1 &= \frac{\partial H}{\partial \pi_1} =  \frac{\left(\sqrt{3}\pi_2 + \pi_1\right)}{2}B \left(\frac{\sqrt{3}q_2 + q_1}{\sqrt{2}}\right) \nonumber \\
		&- \frac{\left(\sqrt{3}\pi_2 - \pi_1\right)}{2}B \left(\frac{\sqrt{3}q_2 - q_1}{\sqrt{2}}\right)  +  2 \pi_1  B(\sqrt{2} q_1), \nonumber\\
		\dot{q}_2 &= \frac{\partial H}{\partial \pi_2} =   \frac{\sqrt{3}\left(\sqrt{3}\pi_2 + \pi_1\right)}{2}B \left(\frac{\sqrt{3}q_2 + q_1}{\sqrt{2}}\right) \nonumber \\ &~~~~~~~~~~ + \frac{\sqrt{3}\left(\sqrt{3}\pi_2 - \pi_1\right)}{2} B \left(\frac{\sqrt{3}q_2 - q_1}{\sqrt{2}}\right), \nonumber\\
		\dot{\pi}_1 &= -\frac{\partial H}{\partial q_1} =   - \frac{\left(\sqrt{3}\pi_2 + \pi_1\right)^2}{4 \sqrt{2}}B' \left(\frac{\sqrt{3}q_2 + q_1}{\sqrt{2}}\right) \nonumber \\
		&- \sqrt{2} \pi_1^2 B'(\sqrt{2} q_1)  + \frac{\left(\sqrt{3}\pi_2 - \pi_1\right)^2}{4 \sqrt{2}}B' \left(\frac{\sqrt{3}q_2 - q_1}{\sqrt{2}}\right) \nonumber, \\
		\dot{\pi}_2 &= -\frac{\partial H}{\partial q_2} =   - \frac{\sqrt{3} \left(\sqrt{3}\pi_2 + \pi_1\right)^2}{4 \sqrt{2}}B' \left(\frac{\sqrt{3}q_2 + q_1}{\sqrt{2}}\right) \nonumber \\
		&~~~~~~~~~~~-\frac{\sqrt{3} \left(\sqrt{3}\pi_2 - \pi_1\right)^2}{4 \sqrt{2}}B' \left(\frac{\sqrt{3}q_2 - q_1}{\sqrt{2}}\right). \label{eq:3_exact_Hamiltons_eom}
	\end{align}
We follow the  general strategy used for the two particle case. Away from the boundaries shown in \cref{fig:Regions_3_label}, we can set $B(\ldots) = 0$ or $1$ (depending on the sector) and $B'(\ldots) = 0$ in \cref{eq:3_exact_Hamiltons_eom} when we have a free trajectory for $q_1$ and $q_2$ while $\pi_1$ and $\pi_2$ are constant. These are stitched together at the boundaries shown in \cref{fig:Regions_3_label} where the momentum jumps discontinuously. We first write down representative piecewise solutions and then evaluate the jump conditions.

	\subsection{Piecewise solutions}
	\label{sec:3_exact_piecewise}
	We first write down solutions for trajectories within the various sectors listed in \cref{fig:Regions_3_label} away from the boundaries. Irrespective of the sector we are considering, $B'(\ldots)$ vanishes and momenta remain constant in time
	\begin{equation}
		\dot{\pi}_\alpha =0 \implies \pi_\alpha(t) = \pi_\alpha(0). \label{eq:3_exact_momentumfrozen}
	\end{equation}
	The evolution of position coordinates $q_\alpha$ takes on the form
        \begin{equation}
            q_\alpha(t) = q_\alpha(0) + \dot{q}_\alpha(0) t.
        \end{equation}
        The velocities $\dot{q}_\alpha(0)$ are determined from the constant momenta $\pi_\alpha(0)$ in different ways depending on the sector. We begin with $\{0_a \ldots 0_f\}$ where all $B(\ldots)$ vanish in \cref{eq:3_exact_Hamiltons_eom} and we have
	\begin{equation}
		\boxed{0_{a-f}:~	\dot{q}_\alpha = 0\implies q_\alpha(t) = q_\alpha(0).}\label{eq:3_exact_0a}
	\end{equation}
	We reproduced the result discussed earlier in \cref{sec:far_sparated} that for particles that are well-separated, the Hamiltonian vanishes, each phase-space coordinate is a constant of motion and dynamics is completely frozen. 
	
	We now consider the sectors $\{1_a, \ldots, 1_f\}$. Due to the symmetry discussed above, these sectors map into each other and it suffices to consider one representative. We choose $1_a$  with no loss of generality. Here, $B(\sqrt{2} q_1) =1$ while $B \left(\frac{\sqrt{3}q_2 \pm q_1}{\sqrt{2}}\right)=0$. This gives us
	\begin{equation}
		\boxed{1_a: \begin{cases}
				\dot{q}_1 = 2 \pi_1 &\implies q_1(t) = q_1(0) + 2 \pi_1(0) t, \\ \label{eq:3_exact_1a}
				\dot{q}_2 = 0~~~ &\implies q_2(t) = q_2(0).
		\end{cases}}
	\end{equation}
	Observe that the trajectories in \cref{eq:3_exact_1a} do not depend at all on the value of $\pi_1(0)$ and there is no motion in the direction where the sector $1_a$ is extended. By rotation, this feature is true for the other sectors $1_b \ldots 1_f$-- motion is entirely along the direction perpendicular to the strip on which the sector is defined and frozen in the parallel direction.
	
	Next, we consider the  sectors $\{2_a, \ldots, 2_f\}$. We choose $2_a$ with no loss of generality. Here, $B \left(\frac{\sqrt{3}q_2 \pm q_1}{\sqrt{2}}\right)=1$ and $B(\sqrt{2} q_1) =0$ and we obtain
	\begin{equation}
		\boxed{2_a: \begin{cases}
				\dot{q}_1 =  \pi_1 &\implies q_1(t) = q_1(0) +  \pi_1(0) t, \\ \label{eq:3_exact_2a}
				\dot{q}_2 = 3\pi_2 &\implies q_2(t) = q_2(0)+3  \pi_2(0) t.
		\end{cases}}
	\end{equation}
	
	Finally, consider the sector $3$ when all $B(\ldots)=1$ and we obtain
	\begin{equation}
		\boxed{3: \begin{cases}
				\dot{q}_1 =  3\pi_1 &\implies q_1(t) = q_1(0) +  3\pi_1(0) t, \\ \label{eq:3_exact_3}
				\dot{q}_2 = 3\pi_2 &\implies q_2(t) = q_2(0)+3  \pi_2(0) t.
		\end{cases}}
	\end{equation}
	
\subsection{Jump conditions}
The piecewise solutions given by \cref{eq:3_exact_0a,eq:3_exact_1a,eq:3_exact_2a,eq:3_exact_3} apply away from the boundaries shown in \cref{fig:Regions_3_label}. We now discuss what happens at the boundaries. As shown in \cref{fig:Regions_3_label}, there are two types of boundaries -- lines where two sectors meet and points where more than two sectors meet. At these boundaries, we expect the momenta $\pi_1$ and $\pi_2$ to  jump discontinuously. The best way to quantify this jump is to use conservation of energy along with combinations of the momentum eom in \cref{eq:3_exact_Hamiltons_eom} that can be easily handled in an integrated form. We will describe this strategy in detail below for a representative of each type of boundary. 

\subsubsection*{$3-2_a$ boundary}
\begin{figure}[!ht]
	\centering
		\includegraphics[width=0.49\textwidth ]{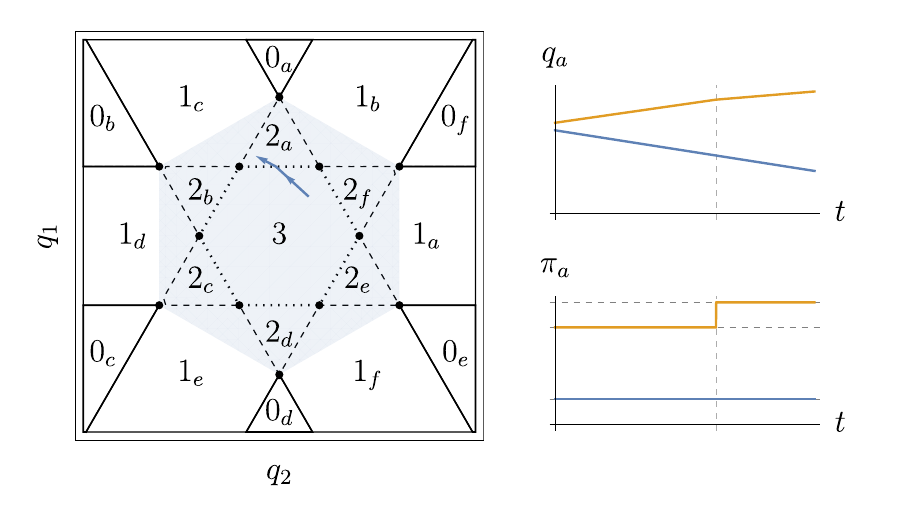}	
	\caption{Numerical plots of a trajectory passing from $3$ to $2_a$ (solid line) obtained for large $\eta$ compared with jump location and magnitudes predicted from the exact solution shown in \cref{eq:jump_3_2a_solved} (broken lines).   \label{fig:3_2_jump}}
\end{figure}	
We begin with the boundary between sectors $3$ and $2_a$ corresponding to the segment
\begin{equation}
 q_1 = \frac{a}{\sqrt{2}},~-\frac{a}{\sqrt{6}}<q_2< \frac{a}{\sqrt{6}}, \label{eq:3_2a_boundary}
\end{equation}
 as shown in \cref{fig:Regions_3_label} and consider the possibility of the trajectory passing this line starting from either side. We want to find how the momenta in the two sectors, $\pi_\alpha^{[3]}$ and $\pi_\alpha^{[2_a]}$ are related to each other. The relationship between momenta and energy in these sectors are
\begin{align}
E{[3]} &= \frac{3}{2}\left( \left(\pi_1^{[3]}\right)^2 + \left(\pi_2^{[3]}\right)^2  \right) , \\
E{[2_a]} &= \frac{1}{2}\left( \left(\pi_1^{[2_a]}\right)^2 + 3\left(\pi_2^{[2_a]}\right)^2  \right) .
\end{align}
Conservation of energy, $E{[2_a]} = E[3]$ giving us one relationship between $\pi_\alpha^{[3]}$ and $\pi_\alpha^{[2_a]}$. A second one can be obtained by considering the equations of motion shown in \cref{eq:3_exact_Hamiltons_eom}. Note that as the trajectory passes through the line separating sectors $3$ and $2_a$, the condition $B' \left(\frac{\sqrt{3}q_2 \pm q_1}{\sqrt{2}}\right) = 0$ remains unchanged, simplifying the effective equations of motion for $\pi_\alpha$
\begin{align}
\dot{\pi}_1 &=   - \sqrt{2} \pi_1^2 B'(\sqrt{2} q_1),~ \dot{\pi}_2 = 0.
\end{align}
This last equation tells us that the $\pi_2$ does not change as the trajectory passes through the boundary and gives us the second relation in addition to the one obtained from conservation of energy. Altogether, we have
\begin{align}
\frac{3}{2}\left( \left(\pi_1^{[3]}\right)^2 + \left(\pi_2^{[3]}\right)^2  \right)&= \frac{1}{2}\left( \left(\pi_1^{[2_a]}\right)^2 + 3\left(\pi_2^{[2_a]}\right)^2  \right) \nonumber,\\
\pi_2^{[3]} &=\pi_2^{[2_a]} \label{eq:jump_3_2a}
\end{align}
Solving these two equations provides us with the jump condition
\begin{equation}
\boxed{\sqrt{3}~\pi_1^{[3]} =\pi_1^{[2_a]}, ~\pi_2^{[3]} =\pi_2^{[2_a]}}. \label{eq:jump_3_2a_solved}
\end{equation}
We see that these solutions exist irrespective of which sector the trajectory is incident from. This means that all trajectories that strike the $3-2_a$ boundary from either sector will pass through. Of course, there exist kinematic constraints that the incident momenta should satisfy to be able to strike the $3-2_a$ boundary. As seen from \cref{fig:3_2_jump} a trajectory incident from the sector $3$ passing through $2_a$ should satisfy $\dot{q_1}^{[3]}>0$ and $\dot{q_1}^{[2_a]}>0$. From \cref{eq:3_exact_3,eq:3_exact_2a}, this translates to conditions on momenta-- $\pi_1^{[3]}>0,\pi_1^{[2_a]}>0$. Similarly, a trajectory incident from the sector $2_a$ passing through $3$ should satisfy  $\dot{q_1}^{[3]}<0$ and $\dot{q_1}^{[2_a]}<0$ which translates to $\pi_1^{[2_a]}<0, \pi_1^{[3]}<0$. \Cref{fig:3_2_jump} shows a sample piece of a trajectory where the above predictions are confirmed against numerical analysis.

\subsubsection*{$2_a-1_b$ boundary}

\begin{figure}[!ht]
	\centering
	\includegraphics[width=0.49\textwidth ]{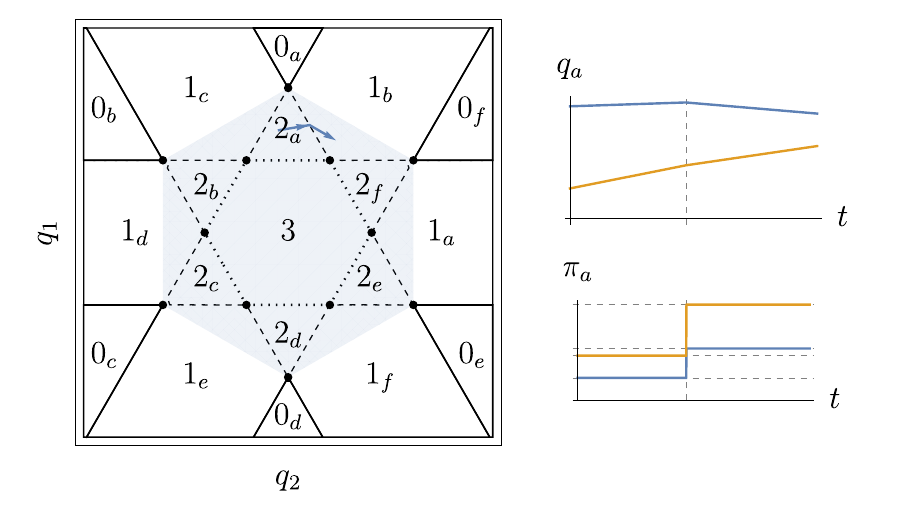}	
	\caption{ Numerical plots of a trajectory passing from $2_a$ to $1_b$ (solid line) obtained for large $\eta$ compared with jump location and magnitudes predicted from the exact solution shown in \cref{eq:jump_2a_1b_solved} (broken lines).   \label{fig:2a_1b_jump}}
\end{figure}	
We  consider now the boundary between sectors $2_a$ and $1_b$ corresponding to the segment  
\begin{equation}
\frac{\sqrt{3} q_2 + q_1}{\sqrt{2}} =  a,~\frac{a}{\sqrt{6}}<q_2< \frac{2a}{\sqrt{6}}, \label{eq:2a_1b_boundary}
\end{equation} shown in \cref{fig:Regions_3_label} and repeat the steps described above. The energy-momentum relation in the two sectors is
\begin{align}
E{[2_a]} &= \frac{1}{2}\left( \left(\pi_1^{[2_a]}\right)^2 + 3\left(\pi_2^{[2_a]}\right)^2  \right), \nonumber\\
E{[1_b]} &=  \frac{1}{4} \left(\sqrt{3}\pi_2^{[1_b]} - \pi_1^{[1_b]}\right)^2.
\end{align}
The effective eom for $\pi_\alpha$ can be obtained by setting $B' \left(\frac{\sqrt{3}q_2 - q_1}{\sqrt{2}}\right) =  B'(\sqrt{2} q_1) = 0$ in \cref{eq:3_exact_Hamiltons_eom} to obtain
\begin{align}
\dot{\pi}_1 &=   - \frac{\left(\sqrt{3}\pi_2 + \pi_1\right)^2}{4 \sqrt{2}}B' \left(\frac{\sqrt{3}q_2 + q_1}{\sqrt{2}}\right),\label{eq:pi1dot_2a_1b}\\
\dot{\pi}_2 & =   - \sqrt{3}  \frac{\left(\sqrt{3}\pi_2 + \pi_1\right)^2}{4 \sqrt{2}}B' \left(\frac{\sqrt{3}q_2 + q_1}{\sqrt{2}}\right).\label{eq:pi2dot_2a_1b}
\end{align}
These two equations can be combined  to obtain
\begin{equation}
\sqrt{3}\dot{\pi}_1 - \dot{\pi}_2 = 0 \label{eq:pidot_2a_1b}
\end{equation}
Imposing energy conservation and integrating \cref{eq:pidot_2a_1b} gives us the following two equations that relates $\pi_\alpha^{[2_a]}$ and $\pi_\alpha^{[1_b]}$
\begin{align}
\frac{1}{2}\left( \left(\pi_1^{[2_a]}\right)^2 + 3\left(\pi_2^{[2_a]}\right)^2 \right)  &= \frac{1}{4} \left(\sqrt{3}\pi_2^{[1_b]} - \pi_1^{[1_b]}\right)^2, \nonumber \\
\sqrt{3} \pi_1^{[2_a]} - \pi_2^{[2_a]} &= \sqrt{3} \pi_1^{[1_b]} - \pi_2^{[1_b]}. \label{eq:jump_2a_1b}
\end{align}
The sector from which the trajectory is incident decides which momenta in \cref{eq:jump_2a_1b} we consider to be the dependent and independent variables. First, we consider trajectories incident from the sector $2_a$ when $\pi_\alpha^{[1_b]}$ should be determined in terms of   $\pi_\alpha^{[2_a]}$. From \cref{fig:2a_1b_jump}, we see that this requires $\dot{q}^{[2_a]}_2 >0$. From the solution shown in \cref{eq:3_exact_2a}, this condition translates to $\pi_2^{[2_a]}>0$. Suppose the trajectory enters $1_b$. The piecewise solution for the trajectory in sector $1_b$ can be determined by rotating \cref{eq:3_exact_1a} to obtain
\begin{align}
 q_1(t) &= q_1(0)  -~ \frac{1}{2} ~\left(\sqrt{3}\pi^{[1_b]}_2 - \pi^{[1_b]}_1\right)t, \nonumber\\ 
 q_2(t) &= q_2(0)  + \frac{\sqrt{3}}{2} \left(\sqrt{3}\pi^{[1_b]}_2 - \pi^{[1_b]}_1\right)t.\label{eq:3_exact_1b}
\end{align}
The physical condition for the trajectory entering $1_b$, $\dot{q}^{[1_b]}_2>0$ gives us the condition $\left(\sqrt{3}\pi^{[1_b]}_2 - \pi^{[1_b]}_1\right) >0$. This picks out the physical solution obtained by solving the quadratic equations in \cref{eq:jump_2a_1b} to give us a unique answer,
\begin{empheq}[box=\fbox]{align}
\pi^{[1_b]}_1 =   \frac{3}{2} \pi_1^{[2_a]} &- \frac{\sqrt{3}}{2} \pi_2^{[2_a]} \nonumber\\&+ \frac{1}{\sqrt{2}}  \sqrt{ \left(\pi_1^{[2_a]}\right)^2 + 3 \left(\pi_2^{[2_a]}\right)^2} \nonumber \\
\pi^{[1_b]}_2 =   \frac{\sqrt{3}}{2} \pi_1^{[2_a]} &-\frac{1}{2}  \pi_2^{[2_a]} \nonumber\\ &+ \sqrt{\frac{3}{2}} \sqrt{ \left(\pi_1^{[2_a]}\right)^2 + 3 \left(\pi_2^{[2_a]}\right)^2}. \label{eq:jump_2a_1b_solved}
\end{empheq}
\Cref{eq:jump_2a_1b_solved} gives us a valid solution for $\pi_\alpha^{[1_b]}$ for any physically allowed incident momenta $\pi_\alpha^{[2_a]}$.  As a result, \emph{all} incident trajectories from $2_a$ that strike the $2_a-1_b$ boundary \cref{eq:2a_1b_boundary} pass on to $1_b$. \Cref{fig:2a_1b_jump} shows a sample trajectory confirming the above results. 

\begin{figure}[!ht]
	\centering
		\begin{tabular}{lr} 
	\includegraphics[width=0.23\textwidth]{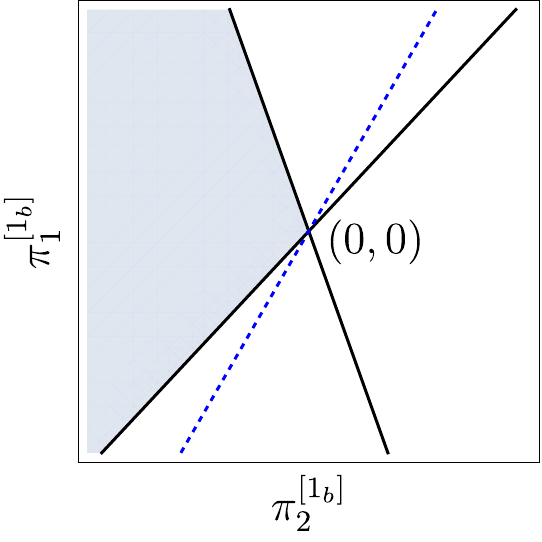}		&
	\includegraphics[width=0.23\textwidth]{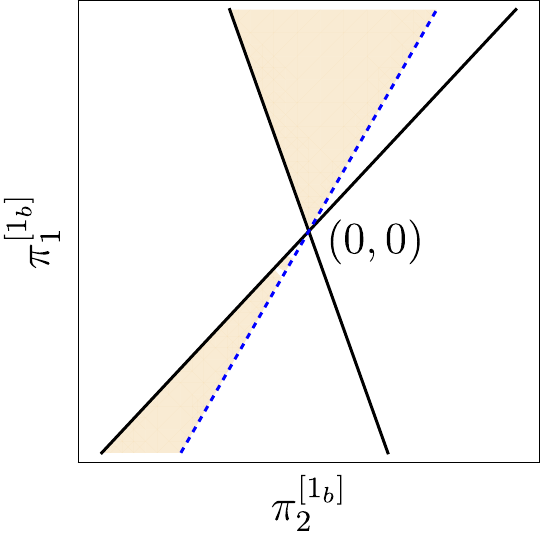}			
		\end{tabular}
	\caption{The union of the shaded regions on the left of the dashed line represents the momenta satisfying \cref{eq:physical_1b_2a} for which trajectories starting from $1_b$ can strike the boundary with $2_a$ (\cref{eq:2a_1b_boundary}). This is divided into momenta satisfying (left) or not satisfying (right) \cref{eq:discriminantge0} for which the trajectories pass on to sector $2_a$ or bounce back into sector $1_b$ respectively. The solid lines are $2\pi_1^{[1_b]} + \sqrt{3} \left( 1 \pm \sqrt{5}\right) \pi_2^{[1_b]} = 0 $ and the dashed line is $\sqrt{3} \pi_2^{[1_b]} = \pi_1^{[1_b]}$. \label{fig:Exact_1b2a}}
\end{figure}

\begin{figure}[!ht]
	\centering
	\includegraphics[width=0.49\textwidth ]{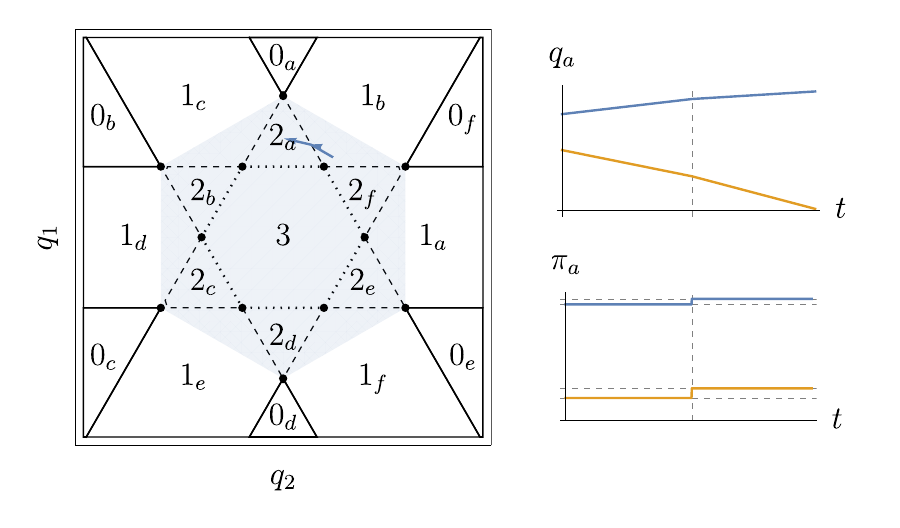}	
	\caption{Numerical plots for large $\eta$ (solid lines) of a trajectory passing from  $1_b$ to $2a$ with initial momenta taking values in the shaded region of the left plot in \cref{fig:Exact_1b2a} compared with jump location and magnitudes predicted from the exact solution shown in \cref{eq:jump_1b_2a_solved} (broken lines).   \label{fig:1b_2a_jump}}
\end{figure}

We now consider the the case when a trajectory is incident from $1_b$. We see from \cref{fig:1b_2a_jump} that such a trajectory requires $\dot{q}^{[1_b]}_2 <0$ which, from  \cref{eq:3_exact_1b} translates to 
\begin{equation}
\left(\sqrt{3}\pi^{[1_b]}_2 - \pi^{[1_b]}_1\right) <0. \label{eq:physical_1b_2a}
\end{equation}
This corresponds to the union of the two shaded regions shown in \cref{fig:Exact_1b2a} lying on the left of the dashed line.
Let us assume that the trajectory passes through into $2_a$, where we require $\dot{q}^{[2_a]}_2 <0$ which imposes $\pi_2^{[2_a]}<0$. We now solve \cref{eq:jump_2a_1b} to obtain $\pi_\alpha^{[2_a]}$ in terms of $\pi_\alpha^{[1_b]}$ subject to these physical conditions to obtain
\begin{empheq}[box=\fbox]{align}
&\pi^{[2_a]}_1 = \frac{9}{10} \pi_1^{[1_b]} - \frac{3 \sqrt{3}}{10}  \pi_2^{[1_b]} \nonumber  \\&-\frac{1}{5} \sqrt{3 \left(\pi_2^{[1_b]}\right)^2 -  \left(\pi_1^{[1_b]}\right)^2 -  \sqrt{3} \pi_1^{[1_b]} \pi_2^{[1_b]}}, \nonumber\\
&\pi^{[2_a]}_2  =- \frac{\sqrt{3}}{10} \pi_1^{[1_b]} + \frac{1}{10}  \pi_2^{[1_b]} \nonumber  \\&-\frac{\sqrt{3}}{5} \sqrt{3 \left(\pi_2^{[1_b]}\right)^2 -  \left(\pi_1^{[1_b]}\right)^2 -  \sqrt{3} \pi_1^{[1_b]} \pi_2^{[1_b]}}. \label{eq:jump_1b_2a_solved} 
\end{empheq}
However, these solutions are not real-valued for all physical values of $\pi_\alpha^{[1_b]}$. Real solutions occur when the discriminant is positive definite
\begin{align}
 \left(\pi_2^{[1_b]}\right)^2 -  \left(\pi_1^{[1_b]}\right)^2 -  \sqrt{3} \pi_1^{[1_b]} \pi_2^{[1_b]} \ge 0. \label{eq:discriminantge0}
\end{align}
This reality condition, combined with \cref{eq:physical_1b_2a} tells us for which incident momenta the trajectory can pass from $1_b$ to $2_a$. This is shown graphically in the left panel of \cref{fig:Exact_1b2a} where the shaded region denotes incident momenta satisfying \cref{eq:discriminantge0,eq:physical_1b_2a}. A sample trajectory satisfying this  is presented in \cref{fig:1b_2a_jump}.
\begin{figure}[!ht]
	\centering
	\includegraphics[width=0.49\textwidth ]{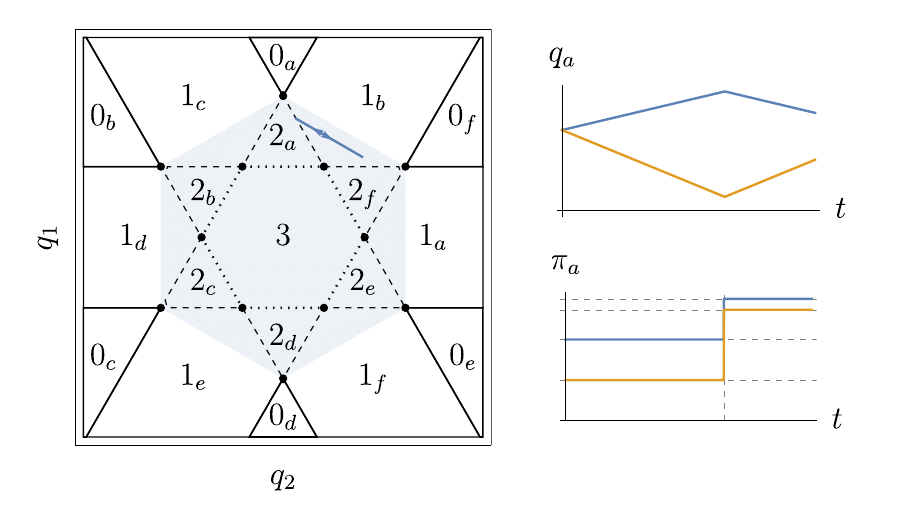}	
	\caption{Numerical plots for large $\eta$ (solid lines) of a trajectory incident from  $1_b$ to the boundary to $2a$ with initial momenta taking values in the shaded region of the right plot in \cref{fig:Exact_1b2a} when the trajectory bounces back to $1_b$. The vertical and horizontal broken lines represent the jump location and magnitudes predicted from the exact solution shown in \cref{eq:jump_1b_1b_solved}.   \label{fig:1b_1b_jump}}
\end{figure}

For other physical values of $\pi_\alpha^{[1_b]}$ satisfying \cref{eq:physical_1b_2a}  but not \cref{eq:discriminantge0} shown in the the shaded region in the right panel of \cref{fig:Exact_1b2a}, the trajectory does not go to sector $2_a$ but `bounces off' to return to $1_b$. We denote by $\pi_{\alpha <}^{[1_b]}$ and $\pi_{\alpha >}^{[1_b]}$ the incident and final momenta, respectively. The latter can be related to the former using energy conservation $E[1_b]_< = E[1_b]_>$ and again using the integrated effective eom \cref{eq:pidot_2a_1b} to obtain the following two equations 
\begin{align}
\frac{1}{4} \left(\sqrt{3}\pi_{2<}^{[1_b]} - \pi_{1<}^{[1_b]}\right)^2 &= \frac{1}{4} \left(\sqrt{3}\pi_{2>}^{[1_b]} - \pi_{1>}^{[1_b]}\right)^2, \nonumber \\
\sqrt{3} \pi_{1<}^{[1_b]} - \pi_{2<}^{[1_b]} &= \sqrt{3} \pi_{1>}^{[1_b]} - \pi_{2>}^{[1_b]}. \label{eq:jump_1b_1b}
\end{align}
A physical requirement is $\dot{q}^{[1_b]}_{2<} <0$ and $\dot{q}^{[1_b]}_{2>} >0$ before and after striking the boundary which gives us
\begin{equation}
\left(\sqrt{3}\pi^{[1_b]}_{2<} - \pi^{[1_b]}_{1<}\right) <0 \text{ and } \left(\sqrt{3}\pi^{[1_b]}_{2>} - \pi^{[1_b]}_{1>}\right) >0. \label{eq:1b1b_bounce_physical}
\end{equation}

Solving \cref{eq:jump_1b_1b} subjected to  \cref{eq:1b1b_bounce_physical} gives us the unique solution 
\begin{equation}
\boxed{\pi^{[1_b]}_{1>} = 2 \pi^{[1_b]}_{1<} - \sqrt{3} \pi^{[1_b]}_{2<},~~\pi^{[1_b]}_{2>} = \sqrt{3} \pi^{[1_b]}_{1<} - 2 \pi^{[1_b]}_{2<}}. \label{eq:jump_1b_1b_solved}
\end{equation}
\Cref{fig:1b_1b_jump} shows such a sample trajectory.

\subsubsection*{$1_a - 0_f$ boundary}

\begin{figure}[!ht]
	\centering
	\includegraphics[width=0.49\textwidth ]{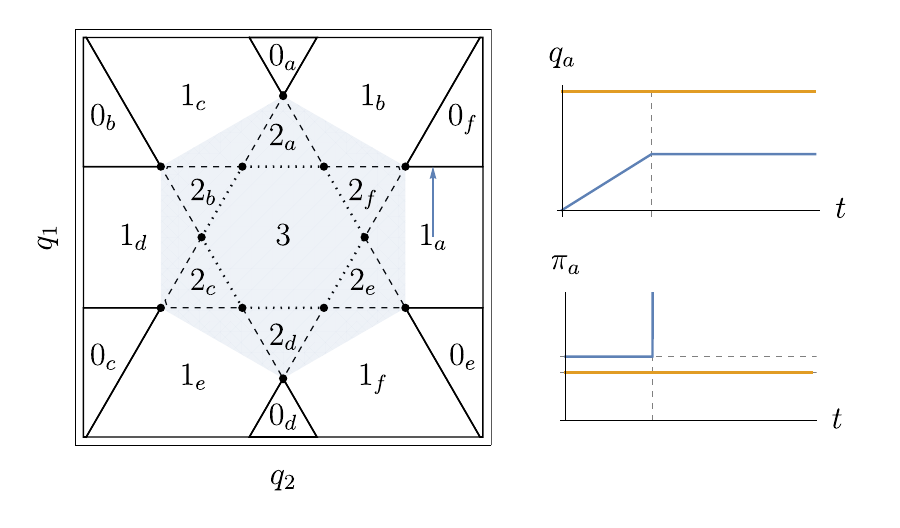}	
	\caption{Numerical plots for large $\eta$ of a trajectory incident from  $1_a$ to the boundary to $0_f$. The vertical and horizontal  broken lines represent the jump location and magnitudes predicted from the exact solution.   \label{fig:1a_0f_jump}}
\end{figure}
We now consider the boundary between sectors $1_a$ and $0_f$ corresponding to the semi-infinite line
\begin{equation}
\sqrt{2}q_1 = a,~ q_2 > \sqrt{\frac{3}{2}}a \label{eq:1a_0f_boundary}.
\end{equation}
The problem is effectively reduced to two-particles considered in \cref{sec:exact_2particles} and can be solved by regulating the system as discussed in \cref{sec:exact_2particles}. The trajectory can strike the boundary \cref{eq:1a_0f_boundary} only if it starts from $1_a$ with $\dot{q}_1>0$ which, from \cref{eq:3_exact_0a} translates to $\pi^{[1_a]}_1 >0$. After striking this boundary, the $q_1$ becomes motionless and we have the jump $\pi_1 \rightarrow \infty$. We have $q_2(t) = q_2(0)$ throughout and it does not change. A sample trajectory is shown in \cref{fig:1a_0f_jump}.

\subsubsection*{$3-2_e-2_f-1_a$ junction}
We now look at the points in \cref{fig:Regions_3_label} where more than two sectors meet beginning with the point 
\begin{equation}
	\left(\tilde{q}_1,\tilde{q}_2\right) = a \left(0,\sqrt{\frac{2}{3}} \right), \label{eq:3_2e_2f_1a_boundary}
\end{equation}
 where sectors $3,2_e,2_f$ and $1_a$ meet. First, we look at the effective eom for $\pi_\alpha$. Since the point $(\tilde{q}_1,\tilde{q}_2)$ is where the lines $\sqrt{3} q_2 \pm q_1 = \sqrt{2} a$ meet, we have $B'\left(\sqrt{2} q_1\right) =0$ and $B'\left(\frac{\sqrt{3} \tilde{q}_2 + \tilde{q}_1}{\sqrt{2}}\right) = B'\left(\frac{\sqrt{3} \tilde{q}_2 - \tilde{q}_1}{\sqrt{2}}\right)$ in the trajectory that goes through this point. Using this in \cref{eq:3_exact_Hamiltons_eom}, we obtain
 \begin{align}
 	\dot{\pi}_1 &=   - \sqrt{\frac{3}{2}} \pi_1 \pi_2~B'\left(\frac{\sqrt{3} \tilde{q}_2 \pm \tilde{q}_1}{\sqrt{2}}\right) \nonumber, \\
 	\dot{\pi}_2 &= - \frac{1}{2} \sqrt{\frac{3}{2}} \left(3 \pi_2^2 + \pi_1^2\right)~ B'\left(\frac{\sqrt{3} \tilde{q}_2 \pm \tilde{q}_1}{\sqrt{2}}\right). 
 \end{align}
This can be massaged to give us
\begin{multline}
\left(3 \pi_2^2 + \pi_1^2\right) \dot{\pi}_1 = 2 \pi_1 \pi_2 \dot{\pi}_2   \implies \frac{d}{dt} \left(\frac{\pi_1^2 + \pi_2^2}{\pi_1^3}\right) = 0. \label{eq:321_triple_eom}
\end{multline}
The integrated version of this equation together with energy conservation  give us the two equations which  can be solved to determine the jump condition for a trajectory that touches the point $(\tilde{q}_1,\tilde{q}_2)$ shown in \cref{eq:3_2e_2f_1a_boundary}. From the solution for the trajectory in $1_a$ shown in \cref{eq:3_exact_1a}, we see that motion only occurs perpendicular to the extended direction in this sector. This does not allow  for the trajectory to either be incident or pass onto $1_a$. Thus, the only trajectories that can strike \cref{eq:3_2e_2f_1a_boundary} are those confined to sectors $3$ or $2_{e,f}$. 

\subsubsection*{$2_a-1_b-1_c-0_a$ junction}
Finally, we look at the point
\begin{equation}
 \left(\tilde{q}_1,\tilde{q}_2\right) = a \left(\sqrt{2},0\right), \label{eq:2a_1b_1c_0a_boundary}
\end{equation}
where sectors $2_a,1_b,1_c$ and $0_a$ meet. Since this  is where the lines $\sqrt{3} q_2 \pm q_1 = \sqrt{2} \pm a$ meet, we have $B'\left(\sqrt{2} q_1\right) =0$ and $B'\left(\frac{\sqrt{3} \tilde{q}_2 + \tilde{q}_1}{\sqrt{2}}\right) = B'\left(\frac{\sqrt{3} \tilde{q}_2 - \tilde{q}_1}{\sqrt{2}}\right)$ for trajectories that strike this point giving us the same condition shown in \cref{eq:321_triple_eom}. This, along with conservation of energy will give us the two equations to determine jump conditions. Since there is no motion in sector  $0_a$,  trajectories striking this point are restricted to $2_a$ and $1_{b-c}$. 

Jump conditions for trajectories striking line-line boundaries required us to solve quadratic equations for which we have presented exact solutions. The jump conditions for trajectories that strike the two classes of point-like boundaries described above requires us to solve cubic equations arising from \cref{eq:321_triple_eom} along with the quadratic equations from the conservation of energy. This too can be done exactly in principle to determine the various possible bounce and pass-through scenarios. We will  not do this explicitly as it would not add much to the discussion.

\subsection{Sample  trajectory and discussion}
\begin{figure}[!ht]
	\centering
	\includegraphics[width=0.26\textwidth]{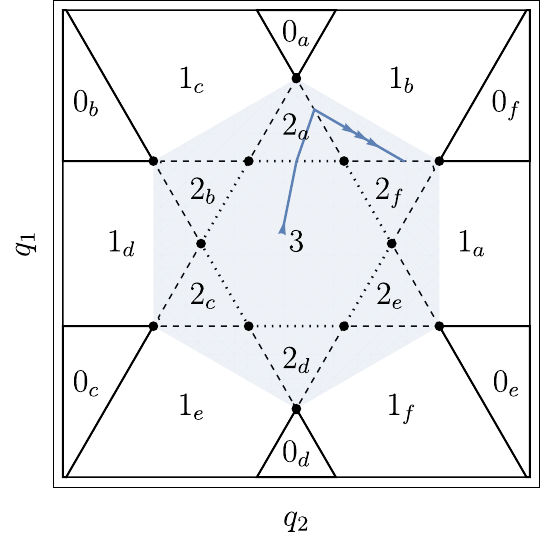}	
	\includegraphics[width=0.46\textwidth]{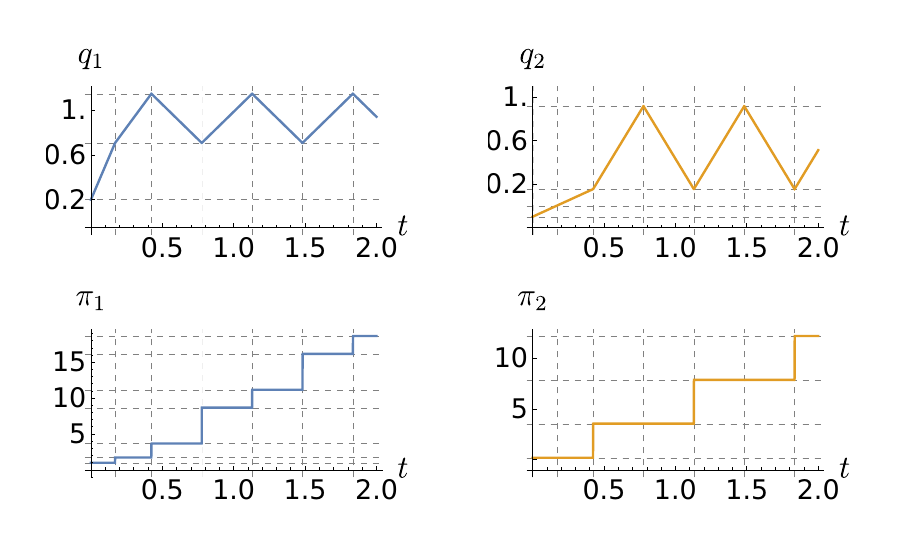}	
	\caption{Numerical plots of a sample trajectory starting from the central sector $3$. The broken lines represent the theoretically determined values between which the full trajectory interpolates.   \label{fig:exact_sample}}
\end{figure}

As an example, we combine all these different behaviors together to understand the numerical trajectory shown in \cref{fig:exact_sample} with thick lines. Here,  the trajectory begins in sector $3$ and settles down to an indefinite oscillation in sector $1_b$, while the momenta increase indefinitely, consistent with the picture in \cref{fig:classical_3particles_fracton}. We mark the various locations where the trajectory crosses sectors along with associated momentum jump values using broken lines.  

From the piecewise solution, we see that for any initial condition  starting anywhere within the shaded region in \cref{fig:Regions_3_label,fig:exact_sample}, the trajectory never leaves the region. And conversely any initial condition located outside the shaded region never enters the shaded region effectively dividing all trajectories. The latter class always ends on the border between some $0_\alpha$ and $1_\beta$ sectors when the position coordinate becomes frozen and momentum diverges. On the other hand, we observe that without fine-tuning, generic initial conditions starting anywhere within the shaded region in \cref{fig:Regions_3_label,fig:exact_sample} end up in such oscillatory steady states in the triangular shaded slivers in any one of the sectors $1_{a-f}$ of the form shown in \cref{fig:exact_sample}. The asymptotic oscillations correspond to the trajectory repeatedly bouncing off the two borders separating some $1_\beta$ sector with adjacent $2_\alpha$ sectors.

\section{Phase space fragmentation and ergodicity breaking}	
	\label{sec:Discussions}	
We now discuss specific aspects of the fracton dynamics listed in \cref{sec:Classical trajectories,sec:exact_trajectories} and also make contact with known quantum fracton phenomenology explored in previous work~\cite{KhemaniHermeleNandkishore_Shattering_PhysRevB.101.174204,SalaRakovskyVerresenKnapPollmann_FragmentationPhysRevX.10.011047,GorantlaLamSeiberg_UVIR_PhysRevB.104.235116,YouMoessner_UVIR_PhysRevB.106.115145,SkinnerPozderac_Thermalization_2023,MorningstayKhemaniHust_Thermalization_PhysRevB.101.214205,Moudgalya_Fragmentation_2022}.

\subsection{Asymptotic conserved quantities and phase space fragmentation}
	\label{sec:Shattering}
	\begin{figure}[!ht]
		\centering
		\begin{tabular}{cc}
				\includegraphics[width=0.23\textwidth]{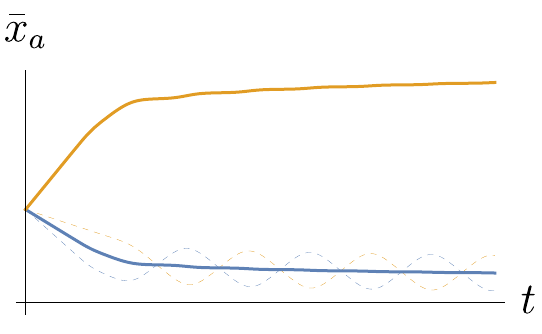} &
				\includegraphics[width=0.23\textwidth]{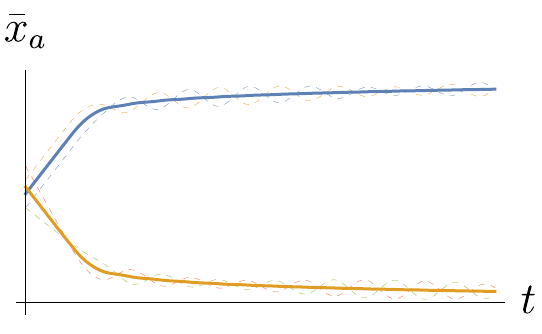} \\
    \includegraphics[width=0.23\textwidth]{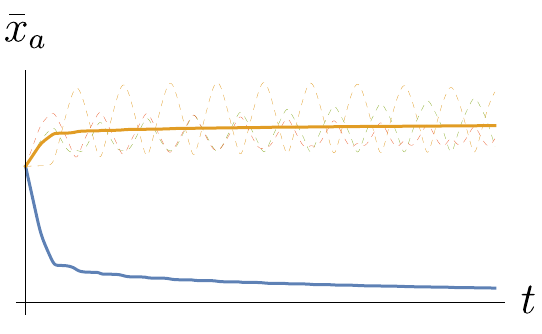} &
    \includegraphics[width=0.23\textwidth]{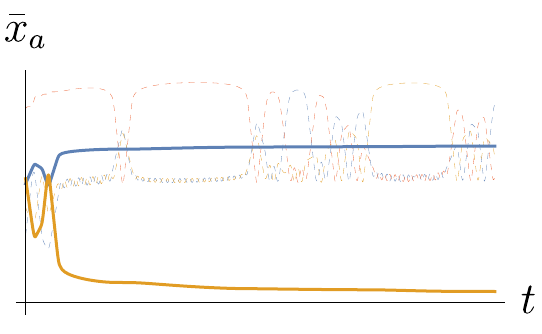} 
		\end{tabular}
		\caption{ Fragmentation of the total center of mass in a single cluster into asymptotically conserved centers of mass for particles in each cluster (thick lines) for the particle trajectories shown in  \cref{fig:classical_3particles_fracton,fig:classical_4particles_fracton} (thin, dashed lines).}
		\label{fig:conserve}
	\end{figure}	
	
	Fractonic trajectories exhibit emergent and asymptotic symmetries and conserved quantities. This was briefly mentioned in passing in \cref{sec:Classical trajectories,sec:exact_trajectories} and we will now discuss it in detail. Emergent conservation laws appear in two distinct ways as a consequence of the interplay between dipole conservation and locality. The first is more direct and determined largely by initial conditions. Recall the discussion in \cref{sec:far_sparated} where we argued that for initially far-separated particles, the dynamics generated by the Hamiltonian in  \cref{eq:Hamiltonian_classical_1d} not only conserves the total dipole moment but also individual particle position coordinates. In other words, the total conserved dipole moment has fragmented into $N$ parts. We will use the symbol $\rightsquigarrow$ to denote the fragmentation of charge as follows
	\begin{equation}
		\sum_a x_a \rightsquigarrow \{x_a\}.
	\end{equation}
        This directly generalizes to the case when particles start off in $P$ well-separated Machian clusters $\cM_1, \ldots, \cM_P$. We expect local dipole conserving dynamics to prevent particles from moving between clusters and as a result, the dipole moment of particles within each Machian cluster $\cM_p$ is independently conserved,
        \begin{equation}
		\sum_a x_a \rightsquigarrow \{\sum_{a \in \cM_p} x_a\}.
	\end{equation}
        A second, less direct way emergent conservation laws appear is when particles start off within a single Machian cluster and asymptotically separate into multiple adjacent clusters. We consider the fractonic trajectories of two particles with close initial separation shown in \cref{fig:classical_2particles}. Here, we see that at late times, not just the total dipole moment of the two particles in the cluster but the coordinate of each individual particle is a conserved quantity 
	\begin{equation}
		x_1 + x_2 \rightsquigarrow \{x_1,x_2\}.
	\end{equation}
	This generalizes to the fractonic trajectories for three and four particles shown in \cref{fig:classical_3particles_fracton,fig:classical_4particles_fracton}. At late times, the particles separate into two subsystems with each undergoing different possible asymptotic dynamics. Regardless of the nature of this dynamics, we see that asymptotically, the dipole moment (center of mass) of each subsystem is independently conserved and becomes constant in time. The precise nature of how the total dipole moment is fragmented depends sensitively on the initial conditions, and can take one of the following forms
	\begin{align}
    &\sum_{a=1}^3 x_a \rightsquigarrow \text{permutations of}~\{x_1,x_2+x_3\} \\
		&\sum_{a=1}^4x_a \rightsquigarrow \text{permutations of}~ \begin{cases}
		    \{x_1 +x_2,x_3+x_4\} \\
      \{x_1,x_2+x_3+x_4\} 
		\end{cases}
	\end{align}
	\Cref{fig:conserve} reproduces the trajectories shown in \cref{fig:classical_3particles_fracton,fig:classical_4particles_fracton} (dashed lines) along with the emergent conserved centers of masses $\bar{x}_a$ (thick lines). For a larger number of particles, we observe a similar phenomenon. However, the precise number of asymptotic clusters is hard to determine because it is hard to conclusively establish that seemingly stable clusters will not interact by exchanging particles at very long timescales. We will not discuss this further in this work. 
	
	\subsection{Ergodicity breaking and the failure of statistical mechanics}
    
	In the study of equilibrium many-body physics, an important role is played by ergodicity and its breaking~\cite{Palmer_BrokenErgodicity}. Unbroken ergodicity allows us to relate physically relevant time-averaged observables,  $\overline{O(\{x_j,p_j\})} $ with calculationally convenient ensemble-averaged ones $\moy{O(\{x_j,p_j\})}$ which is the subject of statistical mechanics. The latter is computed by averaging the observable over all phase space available to the system, $\Gamma$. When ergodicity is broken, the system does not explore all available phase space but gets stuck in distinct sectors $\Gamma \rightsquigarrow \{ \Gamma_\alpha \}$. In certain situations, when $\Gamma_\alpha$ can be well-enumerated, statistical mechanics can still be used by appropriately restricting the ensemble to $\Gamma_\alpha$, using appropriate bias fields or chemical potentials. This is the case when ergodicity breaking occurs through spontaneous symmetry breaking where the sectors $\Gamma_\alpha$ are labelled by elements of the coset $\alpha \in G/H$ of unbroken ($G$) and residual ($H$) symmetries. A dilute gas of dipole conserving systems undergoing fractonic trajectories exhibits ergodicity breaking of a different kind. Here, the nature of the restricted phase space $\Gamma_\alpha$ in which the system becomes trapped depends on the initial conditions and cannot be easily accommodated within statistical mechanics.  
 
 This occurs through two possible mechanisms discussed above in \cref{sec:Shattering}. The first involves particles that start in distant Machian clusters $\cM_\alpha$ that do not merge under dynamics, resulting in the system exploring only a fraction of the symmetry-allowed phase space. The second is the asymptotic fragmentation of particles starting under close separation into multiple sub-clusters as seen in \cref{sec:Classical trajectories,sec:Classical trajectories Unlike charges}. It is natural to postulate that the ergodicity breaking phenomenon survives in a many-body setting for low particle densities when generic initial conditions produce either isolated particles or well-separated clusters of a small number of active particles whose dynamics corresponds to the discussion in \cref{sec:Shattering}. At high densities, we may expect that most initial conditions lead to several particles starting out in a single cluster. In this case, either the system might restore ergodicity beyond some critical density, leading to a thermalization transition as seen in studies of quantum systems~\cite{MorningstayKhemaniHust_Thermalization_PhysRevB.101.214205,SkinnerPozderac_Thermalization_2023} or asymptotic sub-cluster formation persists and presents a barrier to the restoration of ergodicity. In a different work~\cite{AP2023manybodymachian}, we show that the latter scenario is observed with surprising consequences.

\subsection{Ergodicity breaking and Time-reversal symmetry}
\label{sec:timereversal}
\begin{figure}[!ht]
    \centering
    \begin{tabular}{lr}
\includegraphics[width = 0.24\textwidth]{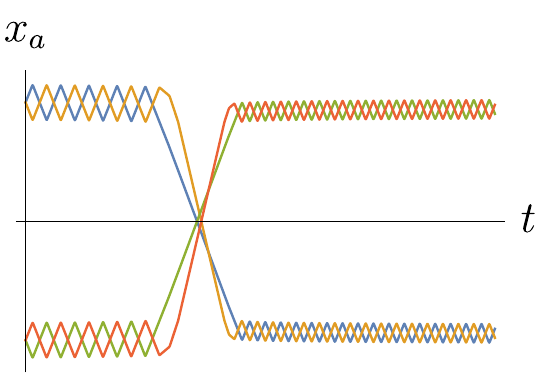}&
    \includegraphics[width = 0.24\textwidth]{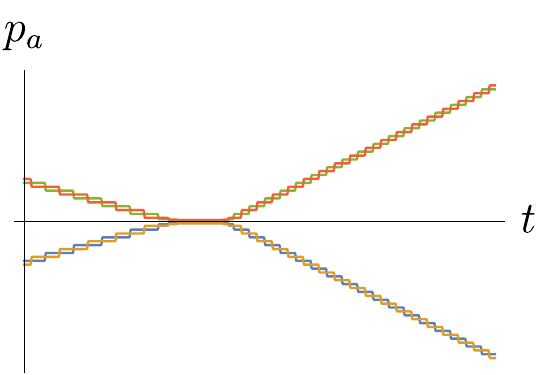}
    \end{tabular}
    \caption{Time-reversed trajectory of four particles with initial conditions when clustering has already set in. We see that the particles briefly reform to a single cluster but eventually break apart at late times. }
    \label{fig:timerev}
\end{figure}

The Hamiltonians we have considered \cref{eq:Hamiltonian_dipole_explicit} has an additional time-reversal symmetry $t \mapsto -t$. It is interesting to understand how this is compatible with cluster formation and ergodicity breaking. Given a typical solution to the equations of motion $\{x_a(t), p_a(t)\}$ for $0\le t \le T$ where particles within a single cluster at $t=0$ have split apart into multiple ones at $t=T$, time-reversal tells us that $\{x_a(T-t), -p_a(T-t)\}$ is a solution to the equations of motion with initial conditions $\{q_a(T), -p_a(T)\}$. In this trajectory, the system starts as multiple clusters at $t=0$ which recombine at $t=T$. However, as illustrated in \cref{fig:timerev} this restoration of ergodicity is only temporary, and if we further evolve the trajectory, the system once again breaks up into clusters. In other words, ergodicity breaking and cluster formation is preserved under time-reversal. 

We contrast this behavior with ordinary classical gases, where ergodicity is \emph{preserved} under time-reversal. A collection of gas particles initially localized spreads out to eventually explore all available phase space, as predicted by ergodicity. However, if we consider its time-reversed trajectory, the particles initially spread out regroup briefly before once again spreading apart.

\subsection{Explicit symmetry breaking and prethermal fracton dynamics}
\label{sec:explicitsymmbreaking}
\begin{figure}[!ht]
    \centering
    \begin{tabular}{lr}
\includegraphics[width = 0.24\textwidth]{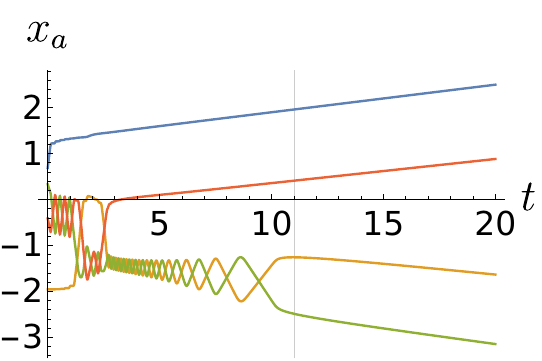}&
    \includegraphics[width = 0.24\textwidth]{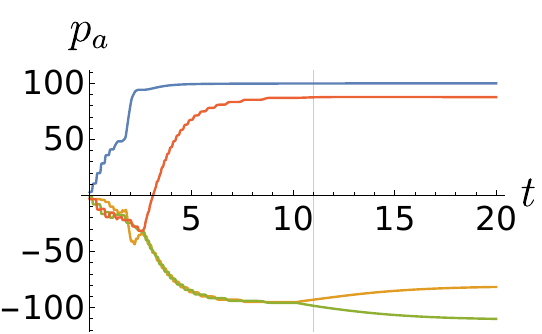}
    \end{tabular}
    \caption{The effect of adding an ordinary kinetic term shown in \cref{eq:symmbreak} with $\epsilon = 10^{-4}$. Fracton dynamics is observed only in an early prethermal regime $t< t_p$ (marked by a vertical line) and eliminated at late times. }
    \label{fig:symmbreak}
\end{figure}

We will now demonstrate that the unusual fractonic dynamics presented so far is a feature of dipole symmetry. To see this, let us consider the effect of explicitly breaking the dipole symmetry by adding the following ordinary kinetic term to the Hamiltonian
\begin{equation}
    \delta H = \epsilon \sum_a p_a^2 \label{eq:symmbreak}.
\end{equation}
As shown in \cref{fig:symmbreak}, the addition of \cref{eq:symmbreak} only preserves fracton dynamics at early times, $t < t_p$ where $t_p$ is a prethermal time-scale set by $\epsilon$. For $t > t_p$ the system behaves like an ordinary system of classical particles: clustering and unbounded momentum growth are eliminated, and ergodicity restored.

    \section{Relation to quantum fractons}
    \label{sec:Quantum}
    For most of this paper, we have used words such as `fractons' without making reference to the quantum mechanical systems and phases of matter whence the name was coined. We now draw comparisons between the phenomena studied in this work and those that are well-known for quantum fractons.

    \subsection{Topological phases, gauging and dipole conservation}
    To keep our work self-contained, we now provide a brief review of quantum fractons and their relation to multipole conserving systems. Readers who are familiar with these can skip ahead to the next subsection. We also direct readers interested in more details to other dedicated reviews~\cite{NandkishoreHermeleFractonsannurev-conmatphys-031218-013604,PretkoChenYou_2020fracton,GromovRadzihovsky2022fractonReview}. 
    
    The term fractons originally referred to quasiparticle excitations in a recently discovered class of gapped phases of quantum matter~\cite{Chamon_Fracton_PhysRevLett.94.040402,Haah_FractonPhysRevA.83.042330,VijayHaahFu_Fractons_PhysRevB.92.235136}. These excitations were found to have restricted mobility and were constrained to move along subdimensional manifolds such as planes, lines~\cite{VijayHaahFu_FractonDuality_PhysRevB.94.235157} or fractals~\cite{Haah_FractonPhysRevA.83.042330}. Fracton phases are `topologically ordered'~\cite{Wen1990topological} in the sense that they are absolutely stable~\cite{CurtVedikaShivaji_AbsolutePhysRevB.94.085112} to \emph{arbitrary} local perturbations~\cite{BravyiHastingsMichalakis_TopologicalOrder_2010}. An important conceptual tool in studying such phases is gauging -- if we start with a system with an unbroken \emph{on-site} global symmetry and dynamically gauge it i.e. elevate it to a local redundancy structure, we obtain a topologically ordered system~\cite{LevinGu_GaugingPhysRevB.86.115109} belonging to the deconfined phase of the gauge theory. It was shown~\cite{VijayHaahFu_FractonDuality_PhysRevB.94.235157,Pretko_FractonGauge_PhysRevB.98.115134,PretkoChenYou_2020fracton,Gromov_Multipole_PhysRevX.9.031035} that  fracton topological order too can be be obtained by starting with systems with an unbroken global multipole conservation symmetry and dynamically gauging it. 
	
    Interestingly, `ungauged' systems with exact global multipole symmetry were found to host a variety of interesting phenomena found in the gauged counter parts and beyond:
    \begin{enumerate}
        \item Their gapped excitations also have restricted mobility~\cite{Pretko_FractonGauge_PhysRevB.98.115134}.
        \item They exhibit Hilbert space fragmentation characterized by a large number of emergent symmetries and conserved quantities~\cite{KhemaniHermeleNandkishore_Shattering_PhysRevB.101.174204,Moudgalya_Fragmentation_2022,SalaRakovskyVerresenKnapPollmann_FragmentationPhysRevX.10.011047}.        
        \item Their dynamics is `glassy' i.e. anomalously slow~\cite{Chamon_Fracton_PhysRevLett.94.040402}.
        \item They exhibit `UV-IR mixing' -- a phenomena where long-distance (IR) behaviour is \emph{not} completely divorced of its microscopic (UV) details~\cite{Minwalla_UVIR_2000,GorantlaLamSeiberg_UVIR_PhysRevB.104.235116,YouMoessner_UVIR_PhysRevB.106.115145}.
    \end{enumerate}
    
     On the one hand, without gauging, these properties are no longer absolutely stable and are conditional on the presence of exact microscopic multipole symmetries. On the other hand, models with exact symmetries are easier to construct and study, and this is the spirit of this work too. 

    \subsection{Classical and quantum fractons - similarities and differences}
    \emph{Restricted mobility:} We begin with the most iconic piece of fracton phenomenology-- particles with restricted mobility. In quantum mechanical models~\cite{VijayHaahFu_Fractons_PhysRevB.92.235136}, excitations of like charges are \emph{completely} immobile whereas composites of charge-neutral dipoles have restricted mobility. We see both of these reflected in the classical dynamics shown in \cref{sec:Classical trajectories}. The latter is evident from the vanishing of the Hamiltonian \cref{eq:Hamiltonian_classical_1d} for well-separated particles, whereas the former is seen in the dynamics of unlike particles shown in \cref{sec:2particles_unlike}. The dynamics of like particles at close separation, shown in \cref{sec:Classical trajectories,sec:exact_trajectories} however, to the best of our knowledge, have no counterpart studied in exactly solvable models~\cite{Haah_FractonPhysRevA.83.042330,VijayHaahFu_Fractons_PhysRevB.92.235136,VijayHaahFu_FractonDuality_PhysRevB.94.235157,Chamon_Fracton_PhysRevLett.94.040402} or in continuum field theories~\cite{GorantlaLamSeiberg_UVIR_PhysRevB.104.235116,Pretko_FractonGauge_PhysRevB.98.115134,Pretko_MachPhysRevD.96.024051}.  The dynamics shown in \cref{sec:Classical trajectories,sec:exact_trajectories} corresponds to the short distance core dynamics of small clusters of particles that are not naturally captured by fixed-point or continuum models. If we consider realistic models with finite correlation length or actual experimental systems, we expect this dynamics (or quantum versions thereof) to be present at low densities of fractons. The family resemblance to the so-called fracton microemulsion phases studied in Ref.~\cite{PremPrtkoNandkishore_FractonPhases_PhysRevB.97.085116} and the bunching/ fracturing instability in Refs.~\cite{BunchingInstability_PhysRevB.107.195131,FBD_PhysRevB.107.195132} are suggestive of this.

    \emph{Hilbert space fragmentation and ergodicity breaking:} In Refs.\cite{KhemaniHermeleNandkishore_Shattering_PhysRevB.101.174204,SalaRakovskyVerresenKnapPollmann_FragmentationPhysRevX.10.011047} it was observed that quantum dynamics preserving dipole moment results in a fragmentation of Hilbert space-- the space of states carrying the same symmetry quantum numbers was found to organize themselves into dynamically disconnected distinct \emph{Krylov sectors}. The size and nature of these Krylov sectors were found to depend on the locality of the Hamiltonian that generates the dynamics. In our classical systems, we discussed that initial conditions separate the particles into distinct Machian clusters that do not merge under dynamics. These disconnected Machian clusters resulting in phase-space fragmentation are direct classical analogues of Krylov subspaces. However, we also saw that particles starting within a single Machian cluster can asymptotically separate into distinct ones resulting in further phase space fragmentation that is not easily determined from initial conditions. To the best of our knowledge, there is no known quantum analog of this asymptotic fragmentation. 
    
    For a low density of particles, the typical size of the Krylov subspace is a vanishing fraction of the total Hilbert space allowed by symmetry and therefore the system violates the eigenstate thermalization hypothesis~\cite{Deutsch_ETH2018} therefore breaking ergodicity, similar to the classical case discussed in \cref{sec:Shattering}.   Recently, Refs. \cite{MorningstayKhemaniHust_Thermalization_PhysRevB.101.214205,SkinnerPozderac_Thermalization_2023} explored an interesting possibility: by tracking the size of the largest Krylov subspace relative to the size of the symmetry sector, the authors showed that the system that breaks ergodicity at low densities can restore it at higher densities, undergoing a sharp transition at a critical density whose critical properties can be determined.  As explained earlier, it is easy to argue that a similar phenomenon of ergodicity breaking also occurs in our system at low densities. In a different work~\cite{AP2023manybodymachian}, we show that, in contrast to the quantum lattice systems of \cite{MorningstayKhemaniHust_Thermalization_PhysRevB.101.214205,SkinnerPozderac_Thermalization_2023}, thermalization does not occur even at high densities in our classical model in the continuum. We remark that although the basic premise of Refs.\cite{MorningstayKhemaniHust_Thermalization_PhysRevB.101.214205,SkinnerPozderac_Thermalization_2023} is quantum mechanical; some analysis is performed using an effective classical Markov system on a lattice. This is still very different from our classical Hamiltonian dynamics.

    	\begin{figure}
		\centering
			\includegraphics[width=0.35\textwidth]{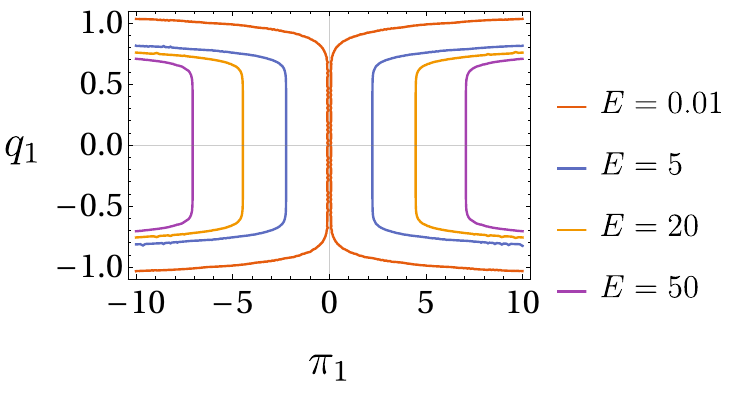} 
		\caption{Energy contours for the two-particle system shown in  \cref{eq:momentum_shell_fracton} with $a=1,\eta=10$. Observe that the hypersurfaces are unbounded in the momentum direction and therefore the usual association between energy and momentum scales no longer holds.  }
		\label{fig:contour}
	\end{figure}	

    \emph{UV-IR mixing:} A general expectation from generic many-body phases is that short-distance (UV) properties of the system on the scale of lattice spacing should not affect long-distance (IR) properties. This allows the use of renormalization group techniques and approximate continuum models for analysis. Long-distance properties of systems with multipole conservation and fracton topological order are known to be sensitive to UV details (such as the lattice structure) and are not easily amenable to analysis using continuum and RG methods. This has been dubbed UV-IR mixing~\cite{Minwalla_UVIR_2000,GorantlaLamSeiberg_UVIR_PhysRevB.104.235116,YouMoessner_UVIR_PhysRevB.106.115145}. One of the ways in which UV-IR mixing manifests itself is in the disassociation between the magnitudes of energy and momenta. We can see classical equivalents of this in the way constant energy hypersurfaces foliate phase space for dipole conserving systems. Recall that for ordinary i.e. dipole non-conserving systems, momentum space projections of constant energy hypersurfaces take the form of spheres whose radius indicates the energy of the system
    \begin{equation}
        \sum_a |\vec{p}_a|^2 \sim 2mE.\label{eq:momentum_shell}
    \end{equation}
    This naturally leads to an association of the momentum scales with energy. For dipole conserving systems on the other hand, this is no longer true. Consider the two-particle dipole conserving system without interactions, written in reduced coordinates.
	\begin{equation}
		E = \pi_1^2 K(\sqrt{2} q_1). \label{eq:momentum_shell_fracton}
	\end{equation}
	Constant energy surfaces of \cref{eq:momentum_shell_fracton} are shown in \cref{fig:contour}. Fractonic trajectories \cref{fig:classical_2particles} appear as open surfaces that are unbounded in the momentum direction. Energy and momentum scales are now no longer related.

	\section{Conclusions and future directions}
	\label{sec:Conclusion}
	In this work, we studied non-relativistic fractons --  point-particle systems with dipole conservation. Using numerical calculations, we show that the classical mechanics of finite number of fractons exhibits several unusual and novel features -- velocity, energy, momentum mismatch, asymptotic steady-state dynamics in the form of attractors and limit cycles  seemingly inconsistent with Liouville's theorem, emergent conserved quantities, and dynamical fragmentation.  We also presented an exact solution for the three-particle system using a certain limiting Hamiltonian form. 
 
 There are several extensions of this work, which involve quantization, many-particles, higher dimensions, higher multipole conservations, and combinations thereof, which are interesting to explore. First, it will be important to extend our analysis to higher dimensions, where more possibilities emerge. The allowed form of the locality term $K$ can take various forms, and different choices can lead to distinct classes of Machian dynamics. It would also be illuminating to canonically quantize the few-particle  system and see to what extent and in what ways if at all the classical features discussed in this work are manifested. Also interesting is to see if a continuum description can be given using our starting point to make contact with field theoretic~\cite{GorantlaLamSeiberg_UVIR_PhysRevB.104.235116} and hydrodynamic studies~\cite{AndrewLucas_2022breakdownofHydro,Gromovetal_FractonHydro_PhysRevResearch.2.033124,LakeClassicalDiffusionhan2023scaling}.  Finally, we would like to identify physical systems where the dynamics described in this work can be potentially observed. One potential setting is systems with strong tilted fields ~\cite{Tilted2021scherg2021observing,Tilted2021morong2021observation,StarkMBL2021PhysRevLett.127.240502} where we can expect a dynamically generated dipole conservation at short times as discussed in \cref{sec:explicitsymmbreaking}. A second possibility is systems with dynamical constraints such as flatbands and Landau levels where dipole symmetry is known to emerge~\cite{DoshiGromov2021vortices,KapustinSpodyneiko_PhysRevB.106.245125,Spodyneiko_GMP_PhysRevB.108.125102}. We leave these and other questions for future work.

	\section*{Acknowledgements}
	  We acknowledge useful discussions with Siddharth Parameswaran,  Michele Fava, Sounak Biswas, Takato Yoshimura, Yuchi He, Dan Arovas, Zohar Nussinov, Yizhi You and Ylias Sadki. We are especially grateful to Fabian Essler for collaboration during the initial stages of this work. We also thank the anonymous referee for useful comments on improving the manuscript.  The work of A.P. was supported by the European Research Council under the European Union Horizon 2020 Research and Innovation Programme, Grant Agreement No. 804213-TMCS  and the Engineering and Physical Sciences Research Council, Grant number EP/S020527/1. The work of A.G. was supported by the Engineering and Physical Sciences Research Council, Grant EP/R020205/1. The work of S.L.S. was supported by a Leverhulme Trust International Professorship, Grant Number LIP-202-014. For the purpose of Open Access, the author has applied a CC BY public copyright license to any Author Accepted Manuscript version arising from this submission.

 \emph{Note added:}
 We became aware of Ref.\cite{Villari2023fracton} which was posted at the same time as our work. This work also considers non-relativistic point-particle fracton systems and studies their linear response theory. Our works are mutually complementary, and therefore there is no overlap, to the best of our knowledge. However, one difference between our formulations is that their Hamiltonian does not include the locality term $K(x)$. It would be interesting to see the effects of its inclusion in their study.

	\bibliography{references}{}

\end{document}